\documentclass{aa}  

\usepackage{amsmath}
\usepackage{amssymb}
\usepackage{graphicx}
\usepackage{color}
\usepackage{txfonts}

\usepackage{xcolor}
\definecolor{xlinkcolor}{cmyk}{1,1,0,0}
 \usepackage[bookmarks=false,         
     pdfnewwindow=true,      
     colorlinks=true,    
     linkcolor=xlinkcolor,     
     citecolor=xlinkcolor,     
     filecolor=xlinkcolor,  
     urlcolor=xlinkcolor,      
final=true
 ]{hyperref}

\usepackage{savesym}
\savesymbol{tablenum}
\usepackage{siunitx}
\usepackage{xspace}
\restoresymbol{SIX}{tablenum}
\usepackage{hyperref}

\newcommand{\St}{\ensuremath{\mathrm{St}}\xspace}

\newcommand\dbquote[1]{\textquotedblleft #1\textquotedblright}
\newcommand\sgquote[1]{\textquoteleft #1\textquoteright}

\newcommand\partialdiff[1]{\frac{\partial}{\partial #1}}
\newcommand\stdiff[1]{\textrm{d} #1}

\newcommand\partiallogdiff[1]{\textrm{d}\mathrm{ln} #1/\textrm{d} \mathrm{ln}\, r}

\definecolor{cgreen}{HTML}{4DAF4A}

\begin{document} 
   \title{Gas accretion damped by dust back-reaction at the snowline}
   \author{Mat\'ias G\'arate\inst{1}
          \and
          Til Birnstiel\inst{1,2}
          \and
          Joanna Dr{\c a}{\.z}kowska\inst{1}
          \and
          Sebastian Markus Stammler\inst{1}
          }
   \institute{$^{1}$University Observatory, Faculty of Physics, Ludwig-Maximilians-Universit\"at M\"unchen, Scheinerstr.\ 1, 81679 Munich, Germany\\
   $^{2}$Exzellenzcluster ORIGINS, Boltzmannstr. 2, D-85748 Garching, Germany\\
              \email{mgarate@usm.lmu.de}
             }

   \date{}

  \abstract
   {The water snowline divides dry and icy solid material in protoplanetary disks, and has been thought to significantly affect planet formation at all stages. 
   If dry particles break up more easily than icy ones, then the snowline causes a traffic jam, because small grains drift inward at lower speeds than larger pebbles.}
   {We aim to evaluate the effect of high dust concentrations around the snowline onto the gas dynamics.}
   {Using numerical simulations, we model the global radial evolution of an axisymmetric protoplanetary disk. Our model includes particle growth, evaporation and recondensation of water, and the back-reaction of dust onto the gas, taking into account the vertical distribution of dust particles.}
   {We find that the dust back-reaction can stop and even reverse the net flux of gas outside the snowline, decreasing the gas accretion rate onto the star to under $50\%$ of its initial value. 
   At the same time the dust accumulates at the snowline, reaching dust-to-gas ratios of $\epsilon \gtrsim 0.8$, and delivers large amounts of water vapor towards the inner disk, as the icy particles cross the snowline.
   However, the accumulation of dust at the snowline and the decrease in the gas accretion rate only take place if the global dust-to-gas ratio is high ($\varepsilon_0 \gtrsim 0.03$), if the viscous turbulence is low ($\alpha_\nu \lesssim \SI{e-3}{} $), if the disk is large enough ($r_c \gtrsim \SI{100}{au}$), and only during the early phases of the disk evolution ($t \lesssim \SI{1}{Myr}$). Otherwise the dust back-reaction fails to perturb the gas motion.}
   {}


   \maketitle
%

\section{Introduction} \label{sec_Intro}
Protoplanetary disks are composed of gas and dust. In the classical picture, a gas disk evolves through viscous evolution driven by outward transport of angular momentum \citep{Lynden-Bell1974}, and orbits at sub-keplerian speed due to its own pressure support.\\
On the other side, dust particles couple to the gas motion according to their size \citep{Nakagawa1986, Takeuchi2002}, small grains quickly follow the motion of the gas, while large boulders are decoupled from it. The mid-sized grains, or pebbles, feel a strong headwind, which causes them to drift towards the gas pressure maximum \citep{Whipple1972, Weidenschilling1977}, which in a typical disk is towards the star.\\
At interstellar dust-to-gas ratios of $1\%$ the force exerted by the dust into the gas is mostly negligible. 
Yet, in regions such as dead zones \citep{Kretke2009, Pinilla2016}, outer edges of gaps carved by planets \citep{Dipierro2017, Kanagawa2018}, snowlines \citep{Brauer2008b, Estrada2016, Drazkowska2017,Stammler2017, Hyodo2019}, and pressure bumps in general \citep{Pinilla2012}, particles can accumulate and grow to larger sizes, reaching concentrations where the dust back-reaction may be strong enough to alter the dynamics of the gas \citep{Taki2016, Onishi2017, Kanagawa2017, Gonzalez2017, Dipierro2018}.\\
In particular, the water snowline acts as a traffic jam for the dust if there is a change in the fragmentation velocity between silicates and ices \citep{Birnstiel2010, Drazkowska2017, Pinilla2017}. 
Previous results showed that the icy particles outside the snowline can grow to larger sizes \citep{Gundlach2011} and drift faster to the inner regions. 
After crossing the snowline, the ice on the solid particles evaporates, leaving only dry silicates behind. Then, the silicates in the inner regions fragment to smaller sizes and drift at lower speeds, creating a traffic jam.
The traffic jam effect can concentrate enough material to trigger the formation of planetesimals through streaming instability \citep{Schoonenberg2017, Drazkowska2017, Drazkowska2018}.\\
In this paper we study the dynamical effect of the snowline on the gas dynamics, by considering the effect of the dust back-reaction onto the gas. We want to find under which conditions the dust can slow down or revert the gas accretion rate, and test if further structures can appear beyond the snowline.\\
We use one-dimensional simulations that consider gas and dust advection, dust growth, and the back-reaction effects. 
To treat the global evolution of the disk we use the model of \citet{Birnstiel2012}, that includes the size evolution of solids by using representative species, and implement the modifications introduced by \citet{Drazkowska2017}, that model the evaporation and recondensation of water at the snowline.\\
The paper is structured as follows. 
In \autoref{sec_GasDustEvolution} we describe the gas and dust velocities considering the back-reaction, and present our model for the snowline.
In \autoref{sec_Setup} we present the setup of our simulations and list the parameter space explored.
In \autoref{sec_Results} we show the conditions in which the accumulation of dust at the snowline results in strong back-reaction effects able to damp the accretion of gas to the inner regions.
In \autoref{sec_Discussion} we discuss the general effects of the back-reaction, when it should be considered, and what observational signatures might reveal dust-gas interactions in the inner regions.
We summarize our results in \autoref{sec_Summary}.
We include a further study of the back-reaction equations and a semi-analytical test for the interested reader in \autoref{Sec_Appendix_EquivalentTest}.
%
\section{Gas and Dust evolution} \label{sec_GasDustEvolution}
The evolution of gas and dust can be described with the advection-diffusion equations as in \cite{Birnstiel2010}:
\begin{equation} \label{eq_gas_advection}
    \partialdiff{t} \left(r \, \Sigma_{\textrm{g}}\right) + \partialdiff{r} (r \, \Sigma_{\textrm{g}} \, v_{\textrm{g,r}})= 0,
\end{equation}
\begin{equation} \label{eq_dust_advection}
    \partialdiff{t} \left(r \, \Sigma_{\textrm{d}}\right) + \partialdiff{r} (r \, \Sigma_{\textrm{d}} \, v_{\textrm{d,r}}) - \partialdiff{r} \left(r D_\textrm{d} \Sigma_\textrm{g} \partialdiff{r}\left(\frac{\Sigma_\textrm{d}}{\Sigma_\textrm{g}}\right)\right)= 0,
\end{equation}
where $r$ is the radial distance to the star, $\Sigma$ is the surface density, $v_\textrm{r}$ is the radial velocity, and $D_\textrm{d}$ is the dust diffusivity. The subindex \sgquote{g} and \sgquote{d} denote the gas and dust, respectively.\\
An expression for the velocities can be obtained from the momentum conservation equations for both components \citep{Nakagawa1986, Tanaka2005, Kanagawa2017, Dipierro2018}, in which the gas feels the stellar gravity, the pressure force, the viscous force, and the drag from multiple dust species, while each dust species only feels the stellar gravity and the drag force from the gas.
%
%
\subsection{Dust Dynamics} \label{sec_dust_dynamics}
Solving the momentum conservation equations, the radial and azimuthal velocities of the dust are as in \cite{Weidenschilling1977, Nakagawa1986, Takeuchi2002}:
\begin{equation} \label{eq_dust_vr}
    v_{\textrm{d},r} = \frac{1}{1 + \mathrm{St}^2} v_{\textrm{g},r} +  \frac{2 \mathrm{St}}{1 + \mathrm{St}^2} \Delta v_{\textrm{g},\theta},
\end{equation}
\begin{equation} \label{eq_dust_vt}
    \Delta v_{\textrm{d},\theta} = \frac{1}{1 + \mathrm{St}^2} \Delta v_{\textrm{g},\theta} - \frac{\mathrm{St}}{2(1 + \mathrm{St}^2)} v_{\textrm{g},r},
\end{equation}
where for convenience the dust azimuthal velocity is written relative to the keplerian velocity $v_K$ as $\Delta v_{\textrm{d},\theta} = v_{\textrm{d},\theta} - v_K$. The same convention is used for the gas azimuthal velocity $\Delta v_{\textrm{g},\theta}$.\\
The Stokes number $\mathrm{St}$ is the dimensionless stopping time that measures the level of coupling of a dust species to the gas motion and is defined as:
\begin{equation} \label{eq_Stokes}
    \mathrm{St}  = t_\textrm{stop}\Omega_K,
\end{equation} 
where $\Omega_K$ is the keplerian angular velocity and $t_\textrm{stop}$ is:
\begin{equation} \label{eq_TimeStop}
    t_\textrm{stop} = \sqrt{\frac{\pi}{8}}\frac{\rho_s}{\rho_\textrm{g}}\frac{a}{c_s},
\end{equation}
with $a$ the particle size, $\rho_s$ the material density of the solids, $\rho_\textrm{g}$ the gas density. The isothermal sound speed $c_s$ is:
\begin{equation} \label{eq_sound speed}
    c_s = \sqrt{\frac{k_\textrm{B} T}{\mu m_\textrm{H}}},
\end{equation}
where $k_\textrm{B}$ is the Boltzmann constant, $T$ the gas temperature, $m_\textrm{H}$ the hydrogen mass, and $\mu$ the mean molecular weight.\\
From \autoref{eq_dust_vr} and \ref{eq_dust_vt} it can be inferred that small particles ($\mathrm{St} \ll 1$) move along with the gas, while large particles  ($\mathrm{St} \gg 1$) are decoupled from it. Particles with $\mathrm{St} \sim 1$ feel the head-wind from the gas with the strongest intensity, and drift most efficiently towards the pressure maximum, in turn, these particles will also exert the strongest back-reaction onto the gas.\\
At the midplane, the Stokes number can be conveniently written as:
\begin{equation} \label{eq_Stokes_Mid}
    \mathrm{St} = \frac{\pi}{2} \frac{a \rho_s}{\Sigma_\textrm{g}}.
\end{equation}
The size of the particles is not static in time \citep{Birnstiel2010, Birnstiel2012}, dust grows until it reaches the fragmentation barrier, where the particles are destroyed by high velocity collisions among themselves \citep{Brauer2008}, or until the drift limit, where they drift faster than they can grow.\\ 
The fragmentation barrier dominates the inner regions of the protoplanetary disk, and the maximum Stokes number that dust grains can reach before fragmenting is:
\begin{equation} \label{eq_frag_limit}
    \mathrm{St}_{\textrm{frag}} = \frac{1}{3}\frac{v_\textrm{frag}^2}{\alpha_t c_s^2},
\end{equation}
where $v_\textrm{frag}$ is the fragmentation velocity which depends on the dust composition, and $\alpha_t$ is the turbulence parameter for the dust fragmentation \citep{Birnstiel2009}.\\
Following \cite{Birnstiel2012}, the drift limit can be approximated by:
\begin{equation} \label{eq_drift_limit}
    \mathrm{St}_{\textrm{drift}} = \left|\frac{\textrm{dln}\, P}{\textrm{dln}\, r }\right|^{-1} \frac{v_K^2}{c_s^2} \epsilon,
\end{equation}
with $v_K$ the Keplerian velocity, and $P$ the isothermal gas pressure at the midplane:
\begin{equation} \label{eq_Pressure_Isothermal}
    P = \frac{\Sigma_\textrm{g}}{\sqrt{2 \pi} h_\textrm{g}} c_s^2,
\end{equation}
with the gas scale height $h_\textrm{g} = c_s/ \Omega_K$.\\
Additionally, we assume that dust diffuses with 
\begin{equation} \label{eq_dust_diffusion}
    D_\textrm{d} = \frac{\nu}{(1+ \epsilon)},
\end{equation}
with $\epsilon = \Sigma_\textrm{d}/\Sigma_\textrm{g}$ the vertically integrated dust-to-gas ratio, and $\nu$ the turbulent viscosity of the gas \citep{Shakura1973}:
\begin{equation} \label{eq_alpha_visc}
\nu = \alpha_\nu \, c_s^{2} \, \Omega_K^{-1},
\end{equation}
controlled by the viscous turbulence parameter $\alpha_\nu$.\\
Notice that as in \cite{Carrera2017}, our model considers two different turbulence parameters: $\alpha_t$ for the dust turbulence (that controls the dust fragmentation, \autoref{eq_frag_limit}), and $\alpha_\nu$ for the viscous turbulence (that controls the gas viscosity, \autoref{eq_alpha_visc}).\\
The $(1+\epsilon)^{-1}$ factor in \autoref{eq_dust_diffusion} comes from considering that the dust concentration diffuses with respect to the gas and dust mixture, instead of the gas only. We neglect the $(1+\mathrm{St}^2)^{-1}$ factor from \cite{Youdin2007} since the particle sizes in our simulations remain small ($\mathrm{St}^2 \ll 1$).
%
\subsection{Gas Dynamics} \label{sec_gas_dynamics}
The gas velocities, considering the dust back-reaction onto the gas, have the following form:
\begin{equation} \label{eq_gas_vr}
    v_{\textrm{g},r} = A v_\nu + 2 B v_P,
\end{equation}
\begin{equation} \label{eq_gas_vt}
    \Delta v_{\textrm{g},\theta} = - A v_P + \frac{1}{2} B v_\nu.
\end{equation}
The gas velocity depends on the viscous velocity $v_\nu$, the pressure velocity $v_P$, and the back-reaction coefficients $A$ and $B$ \citep{Garate2019}.\\
The information related to the dust back-reaction is contained in the coefficients $A$ and $B$, which in a dust free disk have values of $A = 1$ and $B= 0$.\\
In the absence of dust, the gas moves with the viscous velocity \citep{Lynden-Bell1974}:
\begin{equation} \label{eq_visc_velocity}
v_\nu = -\frac{3}{\Sigma_\textrm{g} \sqrt{r}}\partialdiff{r}(\nu \, \Sigma_\textrm{g} \, \sqrt{r}).
\end{equation}
Similarly, if there is no dust, the gas orbits at sub-keplerian speeds due to the pressure support, this pressure velocity is given by:
\begin{equation} \label{eq_press_velocity}
    v_P = -\frac{1}{2} \left( \frac{\Sigma_\textrm{g}}{\sqrt{2\pi}h_\textrm{g}} \, \Omega_K\right)^{-1} \frac{\partial P}{\partial r}.
\end{equation}
The back-reaction coefficients are defined as follows:
\begin{equation} \label{eq_backreaction_A}
    A = \frac{X + 1}{Y^2 +(X+1)^2},
\end{equation}
\begin{equation} \label{eq_backreaction_B}
    B = \frac{Y}{Y^2 +(X+1)^2},
\end{equation}
where $X$ and $Y$ are the following sums defined by \citep{Tanaka2005, Okuzumi2012, Dipierro2018}:
\begin{equation} \label{eq_backreaction_X}
X = \sum_m \frac{1}{1+\mathrm{St}(m)^2} \epsilon(m),
\end{equation}
\begin{equation} \label{eq_backreaction_Y}
Y = \sum_m \frac{\mathrm{St}(m)}{1+\mathrm{St}(m)^2} \epsilon(m),
\end{equation}
where $\epsilon(m)$ is dust-to-gas ratio of the dust species with mass $m$, and the Stokes number can be related to the particle mass through $m = 4\pi a^3 \rho_s/3$ and \autoref{eq_Stokes_Mid}.\\
In principle, \autoref{eq_gas_vr} and \ref{eq_gas_vt} describe gas motion assuming that the dust is well mixed with the gas in the vertical direction (dust-to-gas ratio constant with the distance to the midplane). However, since dust grains settle towards the midplane \citep{Dubrelle1995}, the gas velocity in the midplane layers will be more affected by the dust back-reaction than the surface layers \citep{Kanagawa2017, Dipierro2018}. In section \ref{sec_VerticalApproximation} we show how to calculate the corrected gas velocity, derived from the net mass flux.\\
We discuss a physical interpretation for the back-reaction coefficients $A, B$ in section \ref{sec_interpret_backreaction_coeff}, and provide an approximated expression valid for the case with a single dust species.
\subsubsection{Effect of the vertical structure on the net mass flux} \label{sec_VerticalApproximation}
The corrected gas radial velocity $\bar{v}_\textrm{g,r}$ can be obtained from the net mass flux, which is defined as:
\begin{equation} \label{eq_mass_flux_velocity}
 \Sigma_\textrm{g} \bar{v}_{\textrm{g},r} = \int_{-\infty}^{+\infty} \rho_\textrm{g}(z)\, v_{\textrm{g},r}(z)\, \stdiff{z},
\end{equation}
where $z$ is the distance to the midplane.\\
The vertical profile of the radial velocity $v_{\textrm{g},r}(z)$ depends on:
the vertical density distributions of gas and dust ($\rho_\textrm{g}(z)$, $\rho_\textrm{d}(m, z)$), and the vertical profiles of the viscous and pressure velocities ($v_\nu(z)$, $v_P(z)$).\\
Assuming that the gas and dust are in vertical hydrostatic equilibrium, their respective density profiles are:
\begin{equation} \label{eq_vertical_density_gas}
\rho_\textrm{g}(z) = \frac{\Sigma_\textrm{g}}{\sqrt{2\pi} h_\textrm{g}} \exp\left(-\frac{z^2}{2 h^2_\textrm{g}}\right),
\end{equation}
\begin{equation} \label{eq_vertical_density_dust}
\rho_\textrm{d}(z,m) = \frac{\Sigma_\textrm{d}(m)}{\sqrt{2\pi} h_\textrm{d}(m)} \exp\left(-\frac{z^2}{2 h^2_\textrm{d}(m)}\right),
\end{equation}
where the vertical scale height of the dust species with mass $m$ is defined in \cite{Birnstiel2010} as:
\begin{equation}\label{eq_scaleheight_dust}
h_\textrm{d}(m) = h_\textrm{g} \cdot \min\left(1, \sqrt{\frac{\alpha_t}{\min(\mathrm{St},1/2) (1+\mathrm{St}^2)  }}\right).
\end{equation}
From these profile we can obtain the dust-to-gas ratio of every particle species at height $z$ with:
\begin{equation} \label{eq_dust_to_gas_vertical_generic}
\epsilon(z, m) = \frac{\rho_\textrm{d}(z, m)}{\rho_\textrm{g}(z)},
\end{equation}
and plug it into \autoref{eq_backreaction_X} and \ref{eq_backreaction_Y} to obtain the back-reaction coefficients $A(z)$ and $B(z)$ at every height. At this point we can actually generalize our expression for the gas radial velocity (\autoref{eq_gas_vr}) to every height $v_{\textrm{g},r}(z) = A(z) v_\nu(z) + 2 B(z) v_P(z)$.\\
Now, the final step would be to define the vertical profiles of $v_\nu(z)$ and $v_P(z)$, however we find that assuming $v_\nu (z) = v_\nu$ (as given in \autoref{eq_visc_velocity}), and $v_P (z) = v_P$ (as given in \autoref{eq_press_velocity}) is a good approximation. The net flux is then calculated with \autoref{eq_mass_flux_velocity}.
We test the validity of our approximation in \autoref{sec_Appendix_VerticalApproximation}, and further discuss its physical interpretation.\\
It is worth noting, that under this assumption, the radial gas velocity takes the following form:
\begin{equation}  \label{eq_gas_vr_vertical_average_correction}
 \bar{v}_{\textrm{g},r} =\frac{1}{\Sigma_\textrm{g}} \int_{-\infty}^{+\infty} \rho_\textrm{g}(z)\, 
 \left(A(z)\, v_\nu + 2 B(z)\, v_P\right)\, \stdiff{z} = \bar{A}\, v_\nu + 2 \bar{B}\, v_P,
\end{equation}
where $\bar{A}$ and $\bar{B}$ are the back-reaction coefficients corrected for the vertical structure (i.e., derived from the mass flux in \autoref{eq_mass_flux_velocity}).\\
We repeat the same process to obtain the corrected radial velocity for the dust, by taking the net mass flux for each species of mass $m$:
\begin{equation} \label{eq_mass_flux_velocity_dust}
 \Sigma_\textrm{d}(m) \bar{v}_{\textrm{d},r}(m) = \int_{-\infty}^{+\infty} \rho_\textrm{d}(z, m)\, v_{\textrm{d},r}(z, m)\, \stdiff{z}.
\end{equation}
The gas and dust velocities derived from the net mass flux ($\bar{v}_{\textrm{g},r}$ and $\bar{v}_{\textrm{d},r}$) are used to transport the gas and dust in the advection-diffusion equations \ref{eq_gas_advection} and \ref{eq_dust_advection}.
%
\subsubsection{Understanding the back-reaction coefficients} \label{sec_interpret_backreaction_coeff}
While the back-reaction coefficients may seem rather obscure to interpret at first glance, they can be better understood as a \dbquote{damping} factor (coefficient $A$), that slows the radial viscous evolution and reduces the pressure support, and a \dbquote{pushing} factor (coefficient $B$) that tries to move the gas against the radial pressure gradient and adds some degree of pressure support to the orbital motion.\\
A quick estimate of these coefficients can be obtained if we consider the case of a single (well mixed) particle species \citep{Kanagawa2017, Dipierro2018, Garate2019}:
\begin{equation} \label{eq_backreaction_A_single}
A_\textrm{single} = \frac{\epsilon + 1 + \mathrm{St}^2}{(\epsilon + 1)^2 + \mathrm{St}^2},
\end{equation}
\begin{equation} \label{eq_backreaction_B_single}
B_\textrm{single} = \frac{\epsilon \mathrm{St}}{(\epsilon + 1)^2 + \mathrm{St}^2}.
\end{equation}
From here we can see that both coefficients have values between 0 and 1, and that if the particles are small ($\mathrm{St}^2 \ll \epsilon$), then $A \approx (\epsilon +1)^{-1}$ and $B \approx \mathrm{St}\, \epsilon \,(\epsilon +1)^{-2}$.\\
In the case where the gas velocity $v_\textrm{g,r}$ is dominated by the viscous term such that $A v_\nu > 2B v_P$, the global evolution of gas and dust can be approximated as a damped viscous evolution. In \autoref{Sec_Appendix_EquivalentTest} we further develop this idea, and present a semi-analytical test comparing the evolution of a simulation with back-reaction and dust growth to the standard viscous evolution of a disk with a modified $\alpha_\nu$ parameter.\\
An equivalent expression for the gas and dust velocities, including the contribution of the back-reaction coefficients (\autoref{eq_backreaction_A}, \ref{eq_backreaction_B}), can be found in \cite{Kretke2009}, under the assumption that the particle sizes follow a single power-law distribution.\\
Further analysis on the effects of back-reaction considering different particle size distributions can be found in \cite{Dipierro2018}, where the velocities given in \autoref{eq_gas_vr} and \ref{eq_gas_vt} are equivalent to their Eqs. 11 and 12, while the integrals $X$ and $Y$ are equivalent to their $\lambda_i$ (Eq. 17). \\
%
\subsection{Evaporation and recondensation at the snowline} \label{sec_snowline}
To include the snowline in our simulations, we follow the model given by \citet{Drazkowska2017}, which evolves four different species:
a mix of hydrogen and helium, water vapor, silicate dust, and water ice that freezes over the silicate grains.\\
The gas phase is the sum of both hydrogen-helium and water vapor, it is traced by the surface density $\Sigma_\textrm{g}$, and is advected according to \autoref{eq_gas_advection}.
The water vapor, with surface density $\Sigma_\textrm{vap}$, is advected with the same velocity as the gas, but also diffuses according to the concentration gradient.
The mean molecular weight of the gas phase is then:
\begin{equation} \label{eq_mu_mix}
    \mu = (\Sigma_{\textrm{H}_2} + \Sigma_\textrm{vap}) \left(\frac{\Sigma_{\textrm{H}_2}}{\mu_{\textrm{H}_2}} + \frac{\Sigma_\textrm{vap}}{\mu_\textrm{vap}}\right)^{-1},
\end{equation}
where $\mu_{\textrm{H}_2} = 2.3$ and $\mu_\textrm{vap} = 18$ are respectively the mean molecular weights of the hydrogen-helium mixture and the water vapor, and $\Sigma_{\textrm{H}_2} = \Sigma_\textrm{g}-\Sigma_\textrm{vap}$ is the surface density of the standard hydrogen-helium mixture.\\
The dust grains are assumed to be a mixture of silicates and ices traced by $\Sigma_\textrm{d}$, evolved according to \autoref{eq_dust_advection}, and have a material density of:
\begin{equation} \label{eq_rhos_mix}
    \rho_s = (\Sigma_\textrm{sil} + \Sigma_\textrm{ice}) \left(\frac{\Sigma_\textrm{sil}}{\rho_\textrm{sil}} + \frac{\Sigma_\textrm{ice}}{\rho_\textrm{ice}}\right)^{-1},
\end{equation}
where $\rho_\textrm{sil}$ = $\SI{3}{g\, cm^{-3}}$ and $\rho_\textrm{ice}$ = $\SI{1}{g\, cm^{-3}}$ are the densities of the silicates and ices, respectively, and $\Sigma_\textrm{sil} = \Sigma_\textrm{d} - \Sigma_\textrm{ice}$ is the surface density of the silicates.\\
The composition of the dust grains determines the fragmentation velocity, where icy grains are stickier and can grow to larger sizes than the silicate grains. 
As in \cite{Drazkowska2017}, we assume that the particles have the fragmentation velocity of ices $v_\textrm{frag} = \SI{10}{m\, s^{-1}}$ \citep{Wada2011, Gundlach2011, Gundlach2015} if there is more than $1\%$ of ice in the mixture, and the fragmentation velocity of silicates $v_\textrm{frag} = \SI{1}{m\, s^{-1}}$ \citep{Blum2000, Poppe2000, Guttler2010} otherwise.\\
The limit between evaporation and recondensation of water is given by the equilibrium pressure:
\begin{equation}\label{eq_PressureEq}
    P_\textrm{eq} = P_\textrm{eq,0} \exp(-A/T),
\end{equation}
with $P_\textrm{eq,0} = \SI{1.14e13}{g\, cm^{-1}s^{-2}}$ and  $ A= \SI{6062}{K}$ \citep{Lichtenegger1991, Drazkowska2017}.
The evaporation and recondensation of water are set to maintain the pressure of the water vapor at the equilibrium pressure \citep{Ciesla2006}, with:
\begin{equation} \label{eq_vaporPressure}
    P_\textrm{vap} = \frac{\Sigma_\textrm{vap}}{\sqrt{2\pi}h_\textrm{g}}\frac{k_\textrm{B} T}{\mu_\textrm{vap} m_\textrm{H}}.
\end{equation}
When the water vapor pressure is below this threshold ($P_\textrm{vap} < P_\textrm{eq}$) the ice evaporates into vapor as follows:
\begin{equation}\label{eq_evaporation}
    \Delta \Sigma_\textrm{vap} = \min \left(\sqrt{2\pi} h_\textrm{g} \frac{\mu_\textrm{vap} m_\textrm{H}}{k_\textrm{B} T} (P_\textrm{eq} - P_\textrm{vap}),\, \Sigma_\textrm{ice} \right),
\end{equation}
and vice-versa, if the vapor pressure is higher then it recondenses into ice with:
\begin{equation} \label{eq_recondensation}
    \Delta \Sigma_\textrm{ice} = \min \left(\sqrt{2\pi} h_\textrm{g} \frac{\mu_\textrm{vap} m_\textrm{H}}{k_\textrm{B} T} (P_\textrm{vap} - P_\textrm{eq}),\, \Sigma_\textrm{vap} \right),
\end{equation}
where the factor next to $\pm (P_\textrm{vap} - P_\textrm{eq})$ transforms the pressure difference at the midplane into surface density.\\
As shown by \cite{Birnstiel2010, Drazkowska2017}, at the snowline a traffic jam of dust is created because of the difference in the fragmentation velocities of silicates and ices. Recondensation also contributes to enhance the amount of solids when the vapor diffuses and freezes back beyond the snowline \citep{Stammler2017}. 
%
\section{Simulation Setup} \label{sec_Setup}
We use the code \texttt{twopoppy} \citep{Birnstiel2012} to study the global evolution of a protoplantary disk for $\SI{0.4}{Myr}$ around a solar mass star, advecting the gas and the dust according to the back-reaction velocities described in section \ref{sec_dust_dynamics} and \ref{sec_gas_dynamics}, with the snowline model of \cite{Drazkowska2017} summarized above in section \ref{sec_snowline}.\\
\subsection{Two-Population Dust Model}
In \texttt{twopoppy} the dust is modeled as a single fluid composed of two populations, an initial small particle particle population, and a large particle population with the size limited by the growth barriers (\autoref{eq_frag_limit} and \ref{eq_drift_limit}), with a factor correction: 
$\textrm{St}_\textrm{max} = \min (0.37 \cdot \textrm{St}_{\textrm{frag}}, 0.55 \cdot \textrm{St}_{\textrm{drift}}$).\\
The dust velocity and the back-reaction coefficients are then calculated considering the mass fraction of the two populations. 
\cite{Birnstiel2012} found that the mass fraction of the large population if $f_\textrm{m} = 0.97$ for the drift limited case, and $f_\textrm{m} = 0.75$ for the fragmentation limited case.
\subsection{Disk Initial conditions} \label{sec_SetupDisk}
The gas surface density and temperature profile are defined by the following power laws:
\begin{equation} \label{eq_gas_profile}
    \Sigma_\textrm{g}(r) = \Sigma_0 \left(\frac{r}{r_0}\right)^{-p}, 
\end{equation}
\begin{equation} \label{eq_temperature_profile}
    T(r) = T_0 \left(\frac{r}{r_0}\right)^{-q},
\end{equation}
with $r_0 = \SI{1}{au}$, $\Sigma_0 = \SI{1000}{g\, cm^{-2}}$, $T_0 = \SI{300}{K}$, $p = 1$ and $q = 1/2$.\\ 
The disk surface density initially extends until $r = \SI{300}{au}$.
The disk size is intentionally large to provide a continuous supply of material during the simulation, and to make the interpretation of the back-reaction effects easier. 
We discuss the effect of the disk size in the outcome of the dust accumulation at the snowline in section \ref{sec_Results_LBP}.\\
We start the simulations with an uniform dust-to-gas ratio $\varepsilon_0$ such that $\Sigma_\textrm{d} = \varepsilon_0 \Sigma_\textrm{g}$, assuming that the solid material is composed of a mixture of $50\,\%$ ice and $50\,\%$ silicate \citep[][Table 11]{Lodders2003}. The water vapor is introduced in the simulation as the ice evaporates.\\
The dust phase has a turbulence parameter of $\alpha_t = \SI{e-3}{}$, and an initial size of $a_0 = \SI{1}{\mu m}$.\\
\subsection{Grid and Boundary Conditions}
The region of interest in our simulation extends from $\SI{0.1}{}$ to $\SI{300}{au}$, with $n_r = 482$ logarithmically spaced radial cells.\\
To avoid possible effects of the boundary conditions in our region of interest, we add 20 additional grid cells in the inner region between $\SI{0.05}{}$ and $\SI{0.1}{au}$, and 58 additional grid cells in the outer region between $\SI{300}{}$ and $\SI{600}{au}$. In total, our simulation consist on $560$ grid cells from $\SI{0.05}{}$ to $\SI{600}{au}$.\\
The additional cells at the inner region where added to avoid measuring the accretion rate onto the star too close to the inner boundary. The additional cells in the outer region were added to give the gas enough space to spread outwards without being affected by the outer boundary conditions.\\
At the inner boundary we assume a constant slope for the quantity $\Sigma_\textrm{g,d}\cdot r$. At the outer boundary we have an open boundary condition for the gas and set a constant dust-to-gas ratio (but because of the additional grid cells, the gas never expands all the way to the outer boundary).\\
To calculate the gas and dust velocities and take into account dust settling (\autoref{eq_mass_flux_velocity} and \ref{eq_mass_flux_velocity_dust}) we construct a local vertical grid at every radius with $n_z = 300$ points, logarithmically spaced between $\SI{e-5}{}\, h_\textrm{g}$ and $\SI{10}{}\, h_\textrm{g}$.
\subsection{Parameter Space}\label{sec_ParameterSpace}
\begin{table}
 \caption{Parameter space.}
 \label{TableParam}
 \centering
  \begin{tabular}{ c  c }
    \hline \hline
    \noalign{\smallskip}
    Simulation & $\varepsilon_0$  \\
    \hline
    \noalign{\smallskip}
    Low $\varepsilon_0$ & 0.01  \\
    Mid $\varepsilon_0$ & 0.03  \\
    High $\varepsilon_0$ & 0.05   \\
    \hline
  \end{tabular}
\end{table}
The two most important parameters that control the strength of the back-reaction are the global dust-to-gas ratio $\varepsilon_0$, and the gas viscous turbulence $\alpha_\nu$.\\
We will focus our study in three simulations with \dbquote{Low}, \dbquote{Mid}, and \dbquote{High} global dust-to-gas ratios, with the respective values for $\varepsilon_0$ summarized in \autoref{TableParam}.\\
For the sake of clarity, through the paper we will use a single value for the viscous turbulence, with $\alpha_\nu =\SI{e-3}{}$. This turbulence is low enough for the back-reaction effects to start affecting the gas dynamics (i.e., the term $2\bar{B} v_P$ becomes comparable to $\bar{A} v_\nu$ in the gas velocity, \autoref{eq_gas_vr}).\\
For completeness, in \autoref{sec_Appendix_ParamSpaceExplore} we further extend our parameter space\footnote{The simulation data files, including the extended parameter space, are available in Zenodo: \href{https://doi.org/10.5281/zenodo.3552597}{doi.org/10.5281/zenodo.3552597}} to include different values for the viscous turbulence $\alpha_\nu$, though for simplicity we keep the dust turbulence constant, with $\alpha_t = \SI{e-3}{}$.
%
\section{Dust accumulation and gas depletion at the snowline} \label{sec_Results}
\begin{figure}
\centering
\includegraphics[width=85mm]{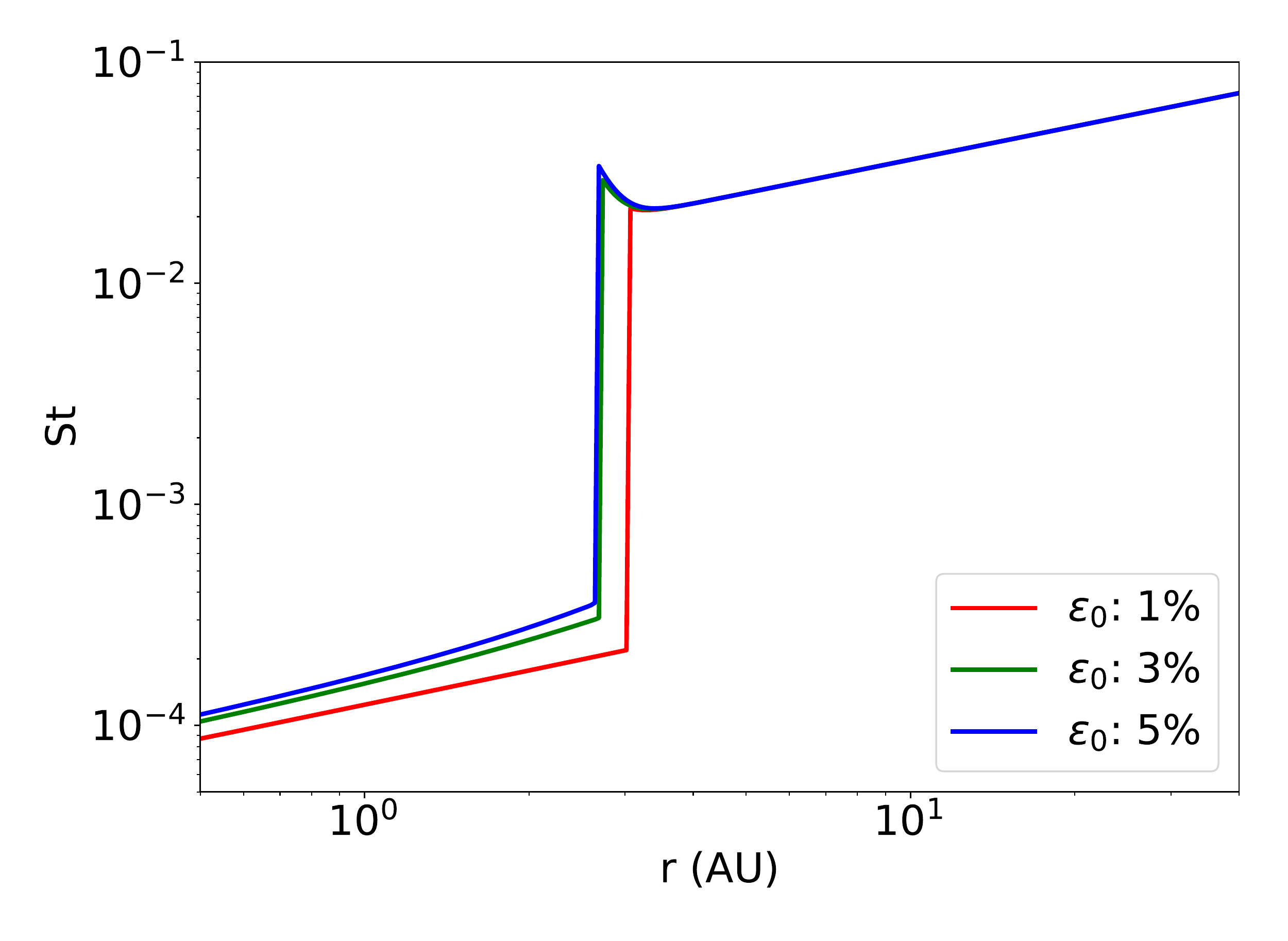}
 \caption{
 Stokes number radial profile after $\SI{0.4}{Myr}$.
 Inside the water snowline (located between $\SI{2.5}{}- \SI{3.0}{au}$) the dust can grow only up to $\mathrm{St} \sim \SI{e-4}{}$. Outside the snowline it can reach values of $\mathrm{St} \sim \SI{e-2}{} - \SI{e-1}{}$.
 }
 \label{Fig_Stokes}
\end{figure}
\begin{figure} 
\centering
\includegraphics[width=85mm]{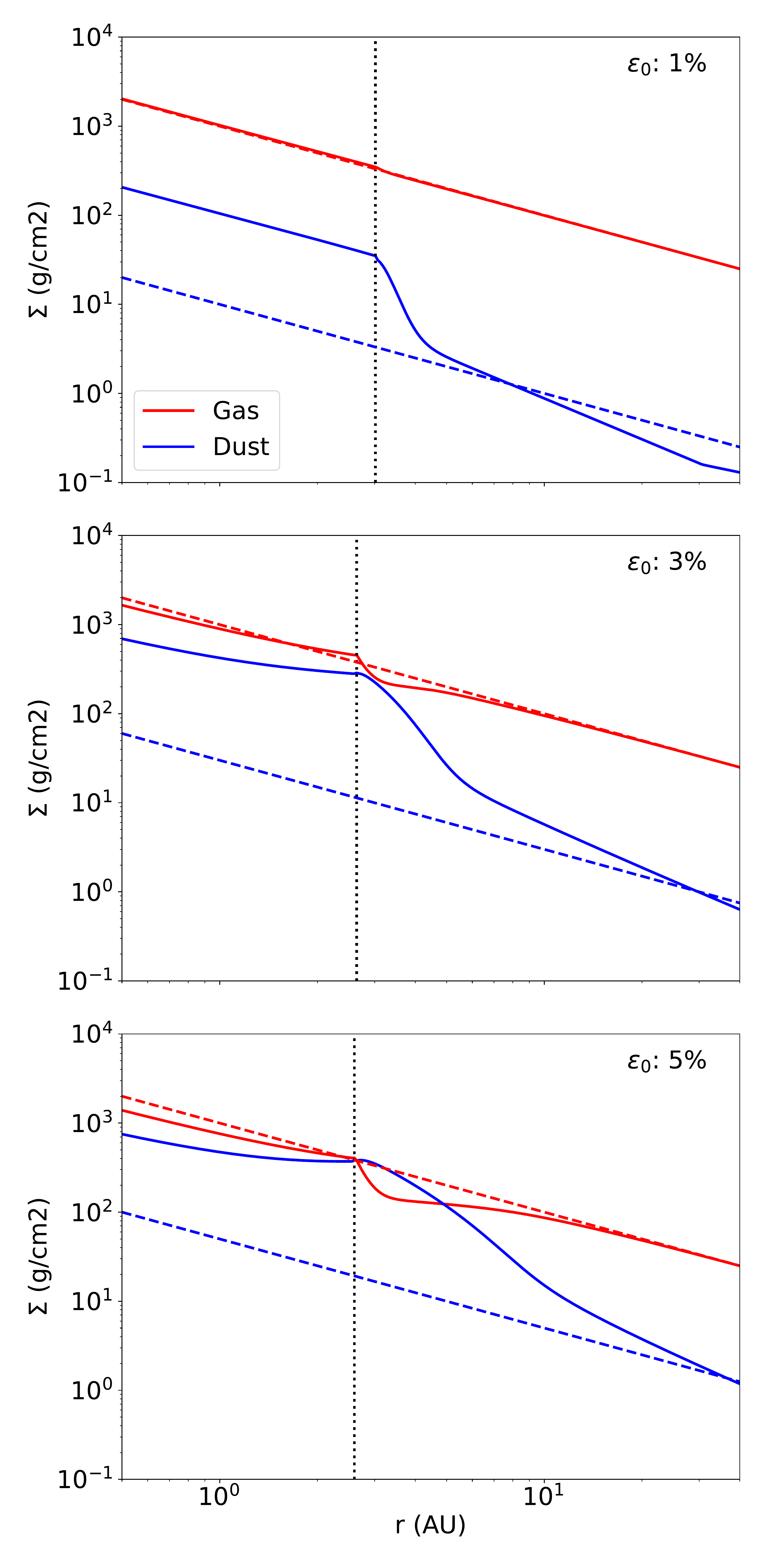}
 \caption{
 Surface density radial profiles of gas (red) and dust (blue) around the snowline.
 The dashed lines mark the initial conditions, and solid lines mark the simulation after $\SI{0.4}{Myr}$. The dotted line marks the snowline at $\SI{0.4}{Myr}$. 
 The \textit{top}, \textit{middle}, and \textit{bottom} panels correspond to the cases with  \dbquote{Low}, \dbquote{Mid}, and \dbquote{High} $\varepsilon_0$, respectively.
 }
 \label{Fig_SurfacePlots}
\end{figure}
\begin{figure} 
\centering
\includegraphics[width=85mm]{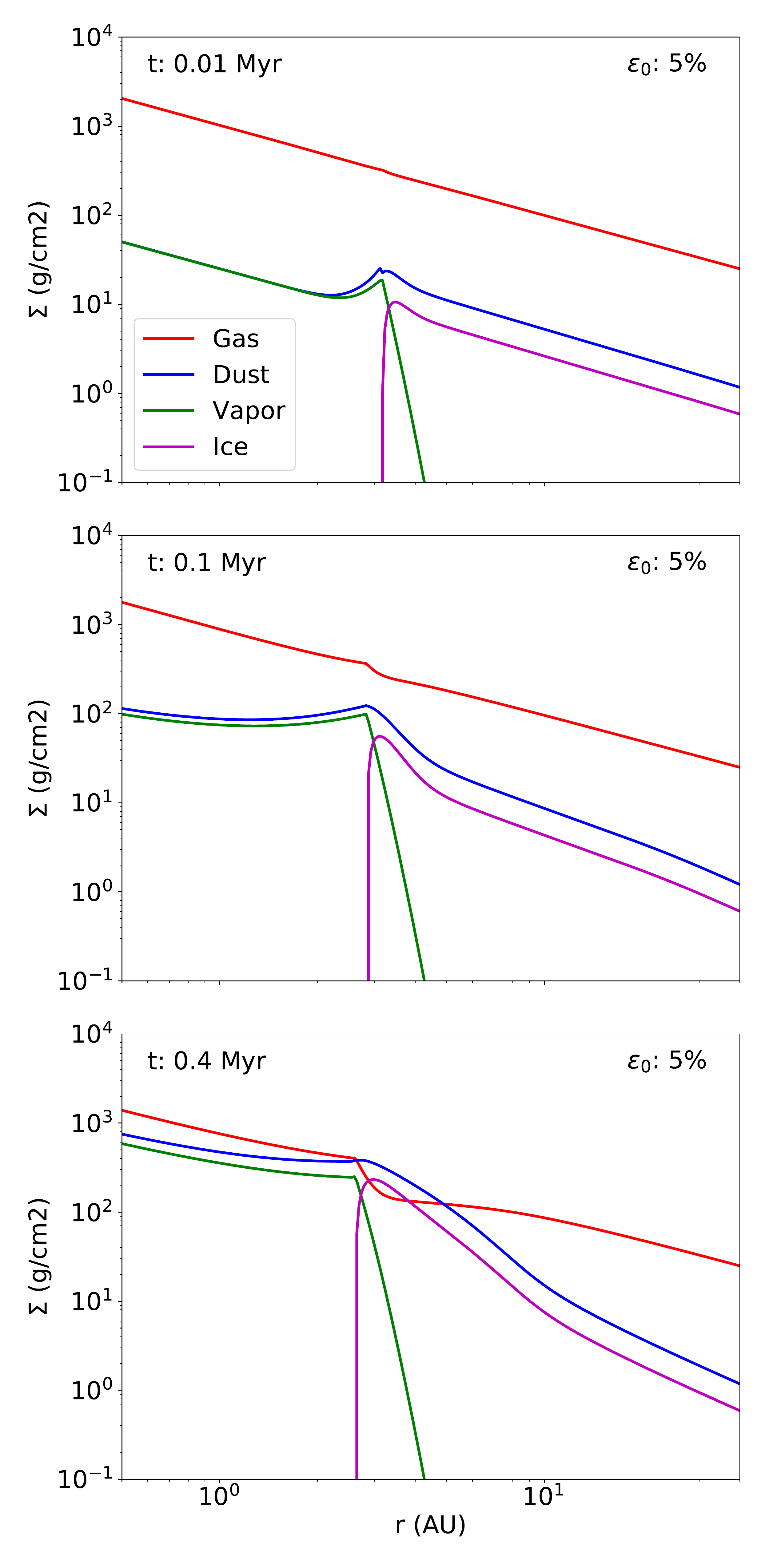}
 \caption{
 Surface densities of gas (red), dust (blue), vapor (green) and ice (purple) of the \dbquote{High $\varepsilon_0$} simulation ($\varepsilon_0 = 0.05$), at different times. 
 As time passes, dust accumulates around the snowline, and the gas surface density is perturbed by the back-reaction.
 }
 \label{Fig_SurfaceEvolution}
\end{figure}
\begin{figure}
\centering
\includegraphics[width=85mm]{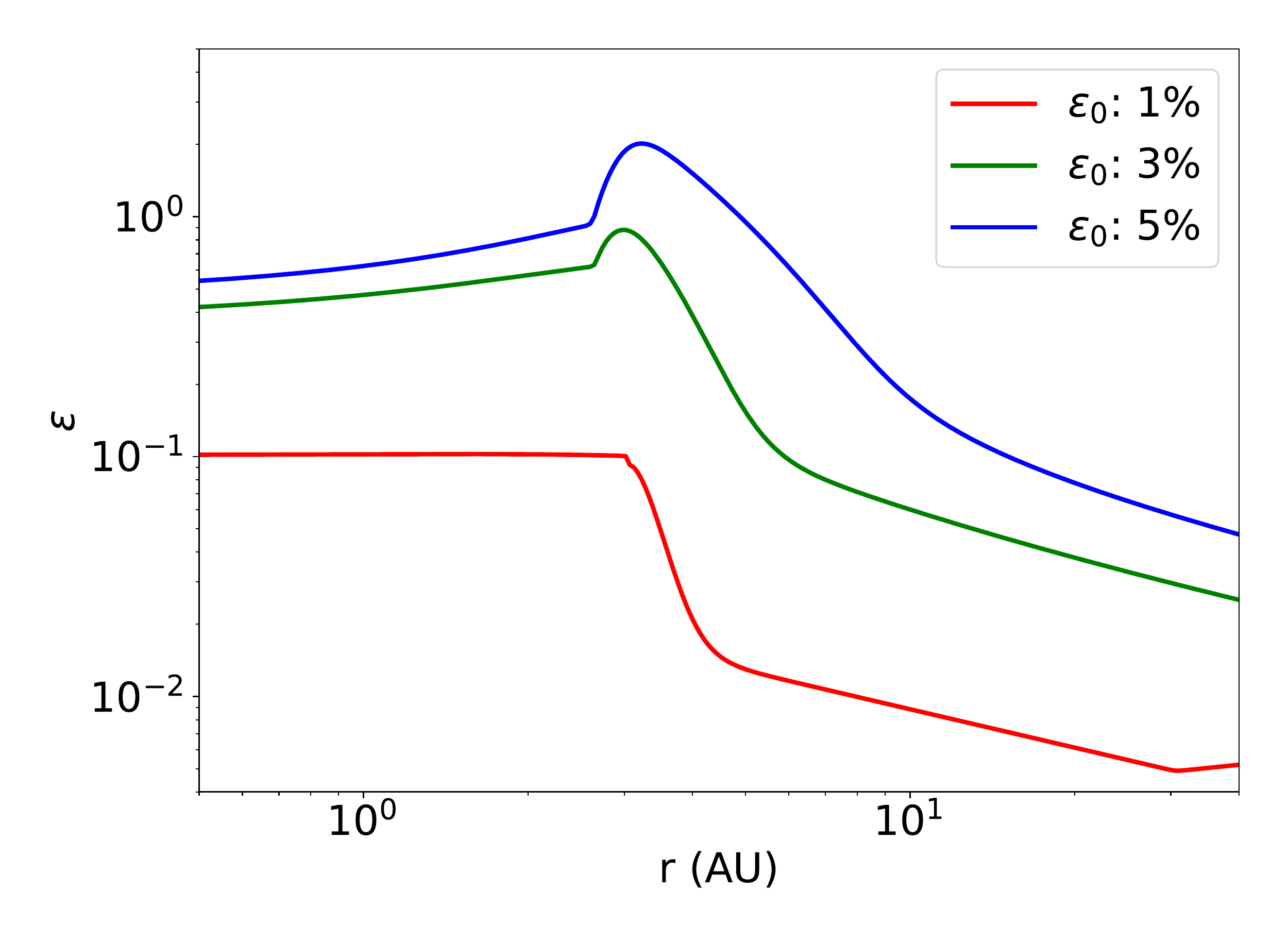}
 \caption{
 Dust-to-gas ratio radial profile for the three simulations after $\SI{0.4}{Myr}$.
 The simulations with a high global dust-to-gas ratio ($\varepsilon_0 \geq 0.03$), shown an enhanced dust accumulation outside the snowline, reaching $\epsilon \approx 0.8\, -\, 2.0 $. 
 }
 \label{Fig_d2gRatio}
\end{figure}
The evolution of gas is initially only dominated by the viscous accretion, but as time passes and dust grows, the back-reaction effects start to become dynamically important to the gas.\\
At the water snowline, the Stokes number changes by 2 orders of magnitude (\autoref{Fig_Stokes}). 
In the inner disk, the particles can only grow to small sizes given by the fragmentation limit of silicates, while in the outer regions the dust size is limited by the fragmentation of water ice or the drift limit.\\
The simulations with the higher dust-to-gas ratio show an increment in the Stokes number at the snowline location, caused by the higher concentration of water vapor which increases the fragmentation limit (by increasing the the mean molecular weight, and decreasing the sound speed, see \autoref{eq_sound speed}, \ref{eq_frag_limit} and \ref{eq_mu_mix}).\\
In the \dbquote{Low $\varepsilon_0$} simulation (\autoref{Fig_SurfacePlots}, top panel), the change in particle size alone causes a traffic jam at the snowline location, as the small dry silicates drift slower than the large icy particles, which results in a higher concentration of dust in the inner regions. 
Outside the snowline the dust-to-gas ratio remains low, so the back-reaction from the large particles is not strong enough to perturb the gas.
In this scenario, the gas surface density remains very close to the initial steady state.\\
Further effects can be seen in the \dbquote{Mid $\varepsilon_0$} simulation (\autoref{Fig_SurfacePlots}, middle panel). First we notice an increment in the gas density profile at the snowline location, caused by the additional water vapor delivered by the icy grains \citep{Ciesla2006}. 
The water vapor and the dust are also more concentrated towards the snowline in this case, as the higher dust-to-gas ratio damps more efficiently the viscous velocity ($|\bar{A} v_\nu| < |v_\nu|$), slowing the diffusion of both gas and small particles. 
At the same time, the additional water vapor also increases the gas pressure, which in turn also increases the drift velocity of the large icy particles towards the snowline, resulting in higher dust concentrations.\\
We also observe a small decrease in the gas surface density outside the snowline, caused by the dust back-reaction that slows down the gas velocity, reducing the supply to the inner regions. This effect becomes more pronounced for higher dust-to-gas ratios.\\
The back-reaction of dust onto the gas causes notorious perturbations in the \dbquote{High $\varepsilon_0$} simulation (\autoref{Fig_SurfacePlots}, bottom panel).
As in the \dbquote{Mid $\varepsilon_0$} simulation, the solids also accumulate at the snowline location, but now the icy dust particles outside the snowline exert a stronger push onto the gas, and reverse the gas accretion of the outer regions. This results in a depletion of gas outside the snowline (between $r > \SI{2.5}{au}$), reaching a minimum density of $\sim 50\%$ of its initial value.\\
Furthermore, the drop in gas density outside the snowline reduces the pressure gradient. Consequently, the drift speed of the large icy particles is also slowed down, allowing for an extended accumulation of dust in the outer regions. 
This process of gas depletion and dust accumulation is expected to continue as long as dust is supplied from the outer regions.\\
In the inner regions inside $\SI{1}{au}$, the gas is depleted to $\sim 65\%$ of its initial value. Only the additional water vapor supplied by the dust crossing the snowline prevents a further depletion of gas.
The evolution of this simulation is illustrated in \autoref{Fig_SurfaceEvolution}, where we can see the initial traffic jam caused by the change in particle size ($t = \SI{0.01}{Myr}$), followed by a further concentration of solids once the vapor accumulates in snowline  ($t = \SI{0.1}{Myr}$), and finally the depletion of gas outside the snowline, accompanied by the extended accumulation of dust  ($t = \SI{0.4}{Myr}$).\\
%
From \autoref{Fig_d2gRatio} we see that the dust-to-gas ratios can reach extremely high values depending on the simulation parameters. 
The \dbquote{Low $\varepsilon_0$} simulation reaches a concentration of $\epsilon \approx 0.1$ in the inner regions  (where the particles are small), because of the traffic jam, but no further accumulation occurs  outside the snowline. \\
In the \dbquote{Mid $\varepsilon_0$} case, the dust-to-gas ratio reaches a high value of $\epsilon\approx 0.85$ at the snowline, and $\epsilon\approx 0.4$ at $\SI{1}{au}$. The dust is more concentrated towards the snowline in this case because the back-reaction slows down the viscous diffusion (\autoref{eq_gas_vr}), yet as time passes the dust should spread more evenly towards the inner regions.\\
The most extreme case is the \dbquote{High $\varepsilon_0$} simulation, where the dust accumulates both inside and outside the snowline.
The dust accumulates in the inner regions due to the traffic jam caused by the change in particle size and the pressure maximum caused by the water vapor, reaching concentrations between $\epsilon \approx 0.5 - 1.0$.
Outside the snowline the dust back-reaction depletes the gas and reduces the pressure gradient, creating another concentration point between $\SI{2.5}{} -  \SI{4}{au}$ where the dust-to-gas ratio reaches values of $\epsilon \approx 1.0 - 2.0$. The recondensation of vapor also contributes to enhance the concentration of solids outside the snowline \citep{Drazkowska2017, Stammler2017}.
%
\subsection{Accretion damped by the back-reaction} \label{sec_accretion_damping}
\begin{figure}
\centering
\includegraphics[width=85mm]{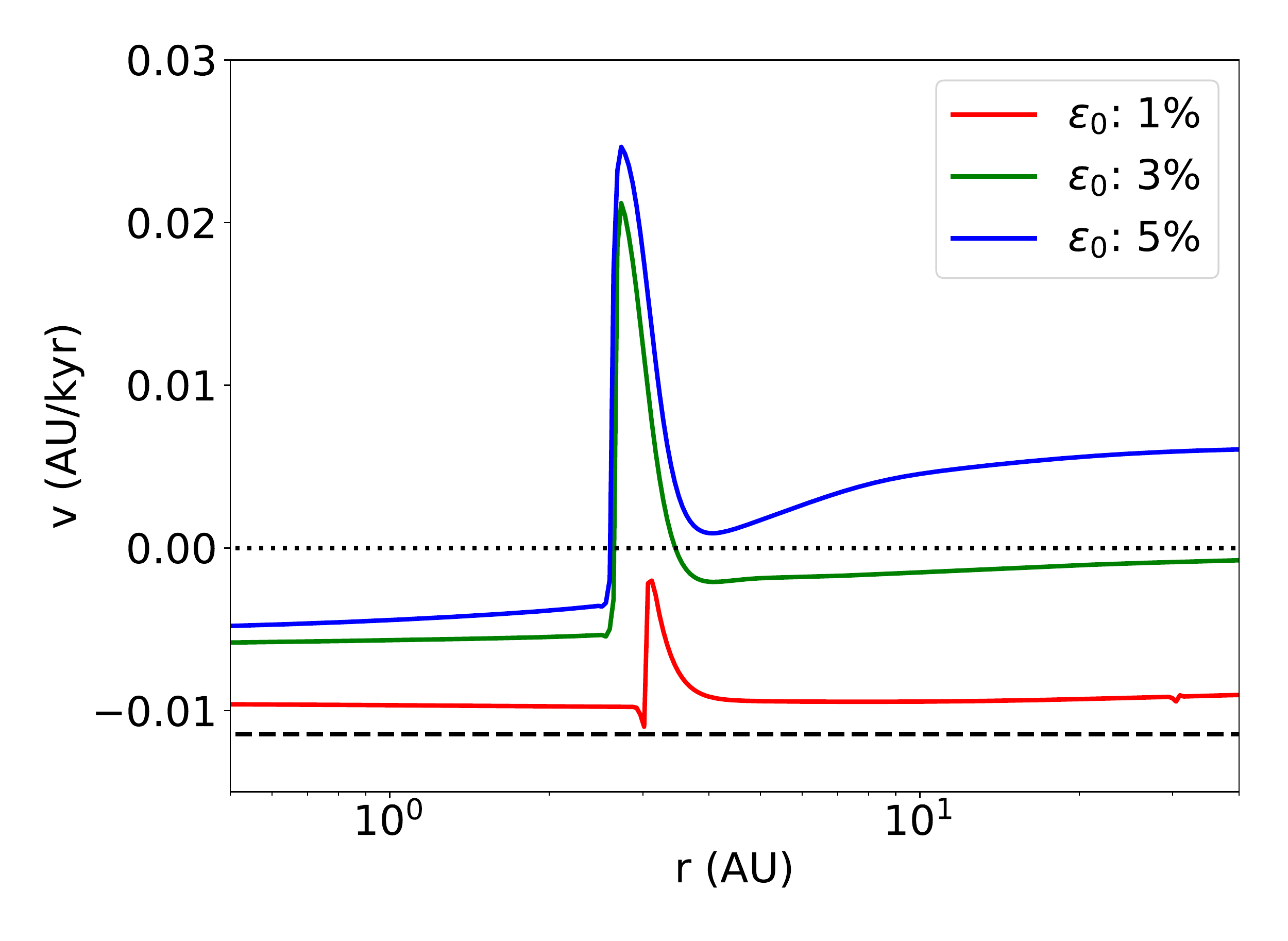}
 \caption{
 Radial gas velocities after $\SI{0.4}{Myr}$ (solid lines), and initial viscous velocity (dashed line).  
 Outside the snowline, the dust back-reaction can stop, and even reverse the gas flux for the simulations with $\varepsilon_0 \geq 0.03$.
 }
 \label{Fig_VelocityDamp}
\end{figure}

\begin{figure}
\centering
\includegraphics[width=85mm]{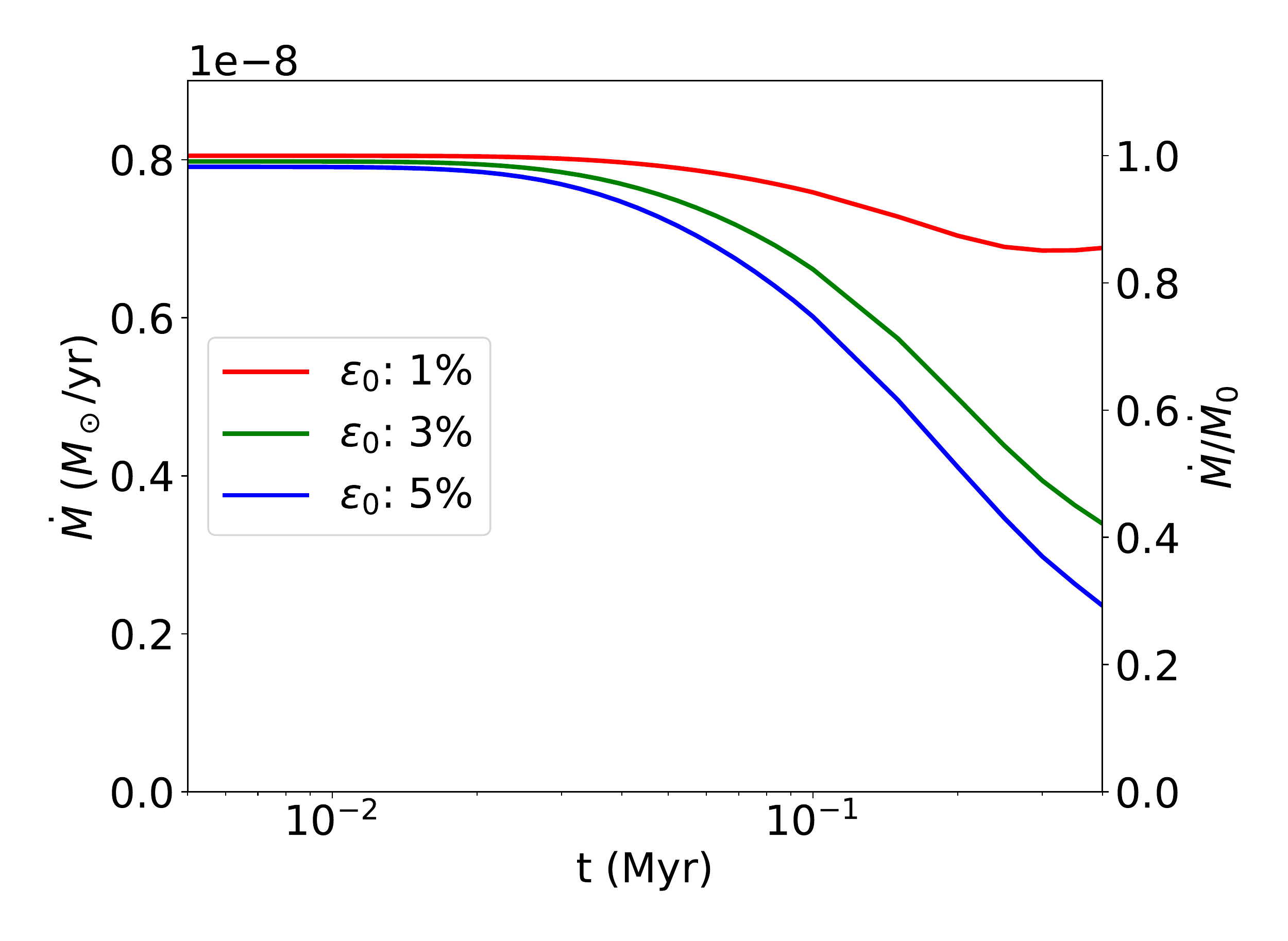}
 \caption{
 Gas accretion rate over time, measured at $\SI{0.5}{au}$. 
  The accretion rate decreases over time, dropping to a $85\%$ of the initial value for the \dbquote{Low $\varepsilon_0$} simulation, and to a $30\% - 45\%$ for the higher $\varepsilon_0$ cases.
 }
 \label{Fig_AccretionDamp}
\end{figure}
\begin{figure}
\centering
\includegraphics[width=85mm]{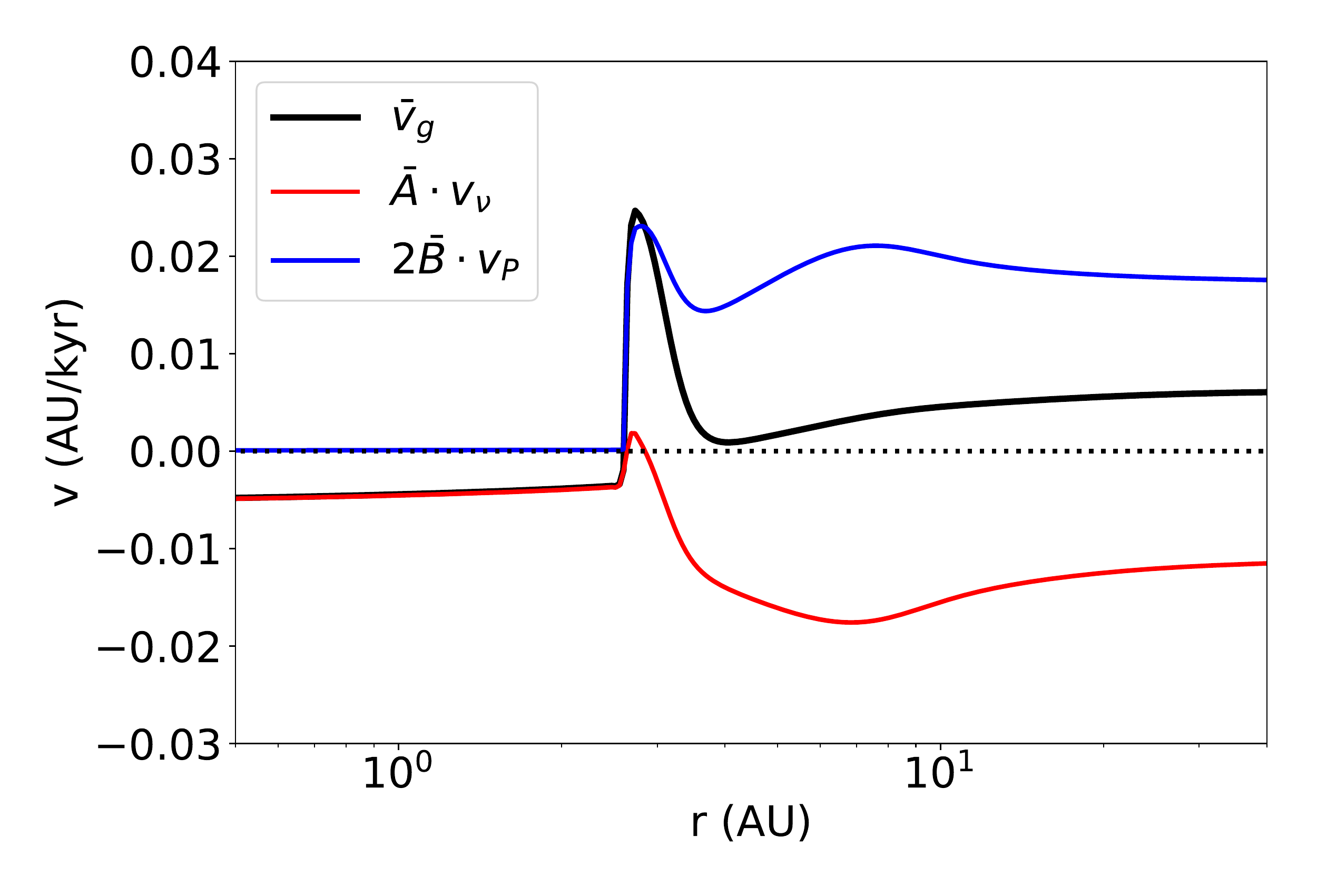}
 \caption{
 Gas velocity profile of the \dbquote{High $\varepsilon_0$} simulation after $\SI{0.4}{Myr}$ (black), and the decomposition of the two velocity terms $\bar{A} v_\nu$ (red) and $2\bar{B} v_P$ (blue) (see \autoref{eq_gas_vr}). 
 In the inner regions the pushing term $2\bar{B} v_P$ is negligible, as the particles Stokes number is too small, and the total velocity is dominated by the damped viscous velocity $\bar{A} v_\nu$.
 In the outer regions the term $2\bar{B} v_P$ overcomes the viscous evolution, and pushes gas against the pressure gradient.
 }
 \label{Fig_VelocityDecomposition}
\end{figure}
The radial velocity of the gas now depends not only on the viscous evolution, but also on the pressure gradient and the dust distribution (\autoref{eq_gas_vr} to \ref{eq_backreaction_Y}). Therefore, for high dust-to-gas ratios and large particles sizes, the gas flow may be damped and even reversed.\\
\autoref{Fig_VelocityDamp} shows the gas velocities of the different simulations.
In the \dbquote{Low $\varepsilon_0$} simulation the dust-to-gas ratio is higher in the inner regions (where grain sizes are small), and lower at the outer regions (where particle sizes are large).
This trade-off between concentration and size means that the dust back-reaction does not dominate the evolution of the gas, and that the gas velocity is only damped with respect to the steady state viscous velocity by a factor of a few.\\
The gas velocity is roughly $ v_\textrm{g,r} \approx 0.85\, v_\nu$ inside the snowline and $v_\textrm{g,r} \approx 0.80\, v_\nu$ outside the snowline, where the transition is caused by the change in both particle size and dust-to-gas ratio.\\
This damping in the viscous velocity also leads into a similar decrease in the gas accretion rate onto the star, from $\Dot{M} = \SI{8e-9}{M_\odot/yr}$ to $\SI{6.8e-9}{M_\odot/yr}$ (\autoref{Fig_AccretionDamp}). 
Once the dust supply is depleted, the accretion rate should return to its steady state value.\\
In the \dbquote{High $\varepsilon_0$} simulation, where the dust concentrations are high inside and outside the snowline, we can see the full effects of dust back-reaction.
In the inner regions ($r< \SI{2.5}{au}$) the particles are small ($\mathrm{St} \sim \SI{e-4}{}$), so the gas velocity is dominated by the term $\bar{A} v_\nu$, which corresponds to the viscous velocity damped by a factor of $\bar{A} \approx (1 + \epsilon)^{-1}$. 
In the outer region ($r > \SI{2.5}{au}$) where the particles are large ($\mathrm{St} \gtrsim \SI{e-2}{}$), the velocity is dominated by the pressure velocity term $2 \bar{B} v_P$, which moves the gas outward, against the pressure gradient (\autoref{eq_gas_vr}).  This reversal of the gas velocity causes the observed depletion in the gas surface density.
\autoref{Fig_VelocityDecomposition} shows the damping and pushing terms of the gas velocity, to illustrate how the gas motion is affected by the dust back-reaction. \\
Since the gas inner disk is disconnected from the outer disk at the snowline in terms of mass transport, the accretion rate into the star is considerably reduced. As solid particles accumulate around the snowline, and the inner regions become more and more depleted of gas, the accretion rate reaches a value as low as $\Dot{M} = \SI{2.5e-9}{M_\odot/yr}$. The only reason why the gas is not further depleted in the inner regions is because of the water vapor delivered by the icy dust particles crossing the snowline \citep{Ciesla2006}. \\
Meanwhile, the mass outside the snowline is transported outwards at a rate of $\sim \SI{e-9}{} - \SI{e-8}{M_\odot/yr}$. No instabilities seem to appear in gas surface density in the outer regions, as the mass transported to the outer disk is only a small fraction of the total disk mass. 
Once the dust supply is exhausted the back-reaction push will stop being effective, and the gas accretion rate should retake the standard viscous evolution.\\
The behavior of the \dbquote{Mid $\varepsilon_0$} simulation  is consistently in between the \dbquote{Low $\varepsilon_0$} and \dbquote{High $\varepsilon_0$} cases, with that the gas flux is practically frozen ($v_\textrm{g,r} \approx 0$) in the outer regions ($r >\SI{4}{au}$).
All simulations show that the back-reaction push is particularly strong in a narrow region outside the snowline (between $r \approx \SI{2.5}{}- \SI{4}{au}$), where the concentration of icy particles increases because of the recondensation of water vapor.\\
In \autoref{sec_LayeredAccretion} we comment on the effects of dust settling on the accretion rate at different heights.
%
%
\subsection{Depletion of $\textrm{H}_2$ and $\textrm{He}$ inside the snowline.} \label{sec_depletionH2}
\begin{figure}
\centering
\includegraphics[width=85mm]{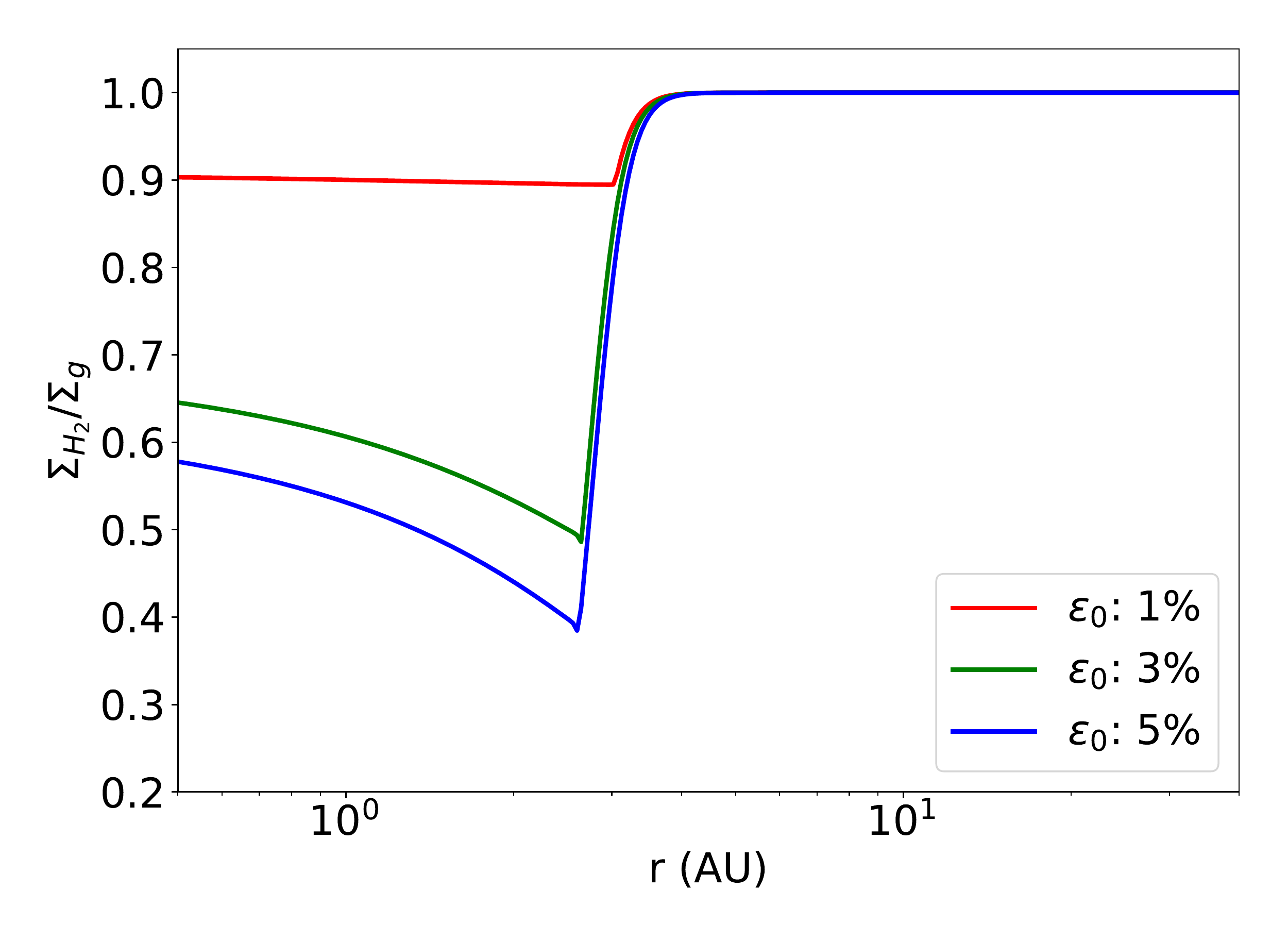}
 \caption{
 $\textrm{H}_2, \textrm{He}$ mass fraction profile after $\SI{0.4}{Myr}$. The mass fraction of light gases is lower inside the snowline as the dust crossing the snowline delivers water vapor. As the global dust-to-gas ratio increases, the  back-reaction push outside the snowline reduces the flux of $\textrm{H}_2, \textrm{He}$ into the inner regions.}
 \label{Fig_H2Abundance}
\end{figure}
From the gas velocities, we see that in the cases where the back-reaction is effective it can stop or reverse the accretion of gas outside the snowline, causing the inner regions to become relatively depleted of gas.\\
In particular, the dust back-reaction reduces the supply of the $\textrm{H}_2, \textrm{He}$ to the inner regions, as outside the snowline this is the dominant gas component.\\
At the same time, the icy grains cross the snowline and deliver water vapor to the inner regions. Therefore, the gas will present a lower $\textrm{H}_2, \textrm{He}$ mass fraction in the inner disk than in the outer disk.\\
The total amount of water delivered to the inner regions depends on the initial dust-to-gas ratio $\epsilon_0$, while the dust back-reaction affects how it is distributed.\\
\autoref{Fig_H2Abundance} shows that even in the \dbquote{Low $\varepsilon_0$} case, the mass fraction of $\textrm{H}_2, \textrm{He}$ is reduced to a $90\%$.\\
For the \dbquote{Mid $\varepsilon_0$} and \dbquote{High $\varepsilon_0$} cases the dust back-reaction onto the gas reduces the supply of light gases to the inner regions, creating environments dominated by water vapor inside the snowline, with a $\textrm{H}_2, \textrm{He}$ mass fraction between $40\% - 65\%$. 
The depletion is more concentrated towards the snowline because the damping term of the gas velocity ($\bar{A}\,v_\nu$) slows down the viscous diffusion of water vapor.\\
After the dust supply is exhausted, the region inside the snowline will be gradually refilled with gas from the outer regions in the viscous timescale ($t_\nu \approx \SI{0.5}{Myr}$ at $\SI{4}{au}$), and the $\textrm{H}_2, \textrm{He}$ mixture will be replenished to become the dominant component once more.
%
\subsection{What happens without the back-reaction?} \label{sec_NoBackReaction}
\begin{figure}
\centering
\includegraphics[width=85mm]{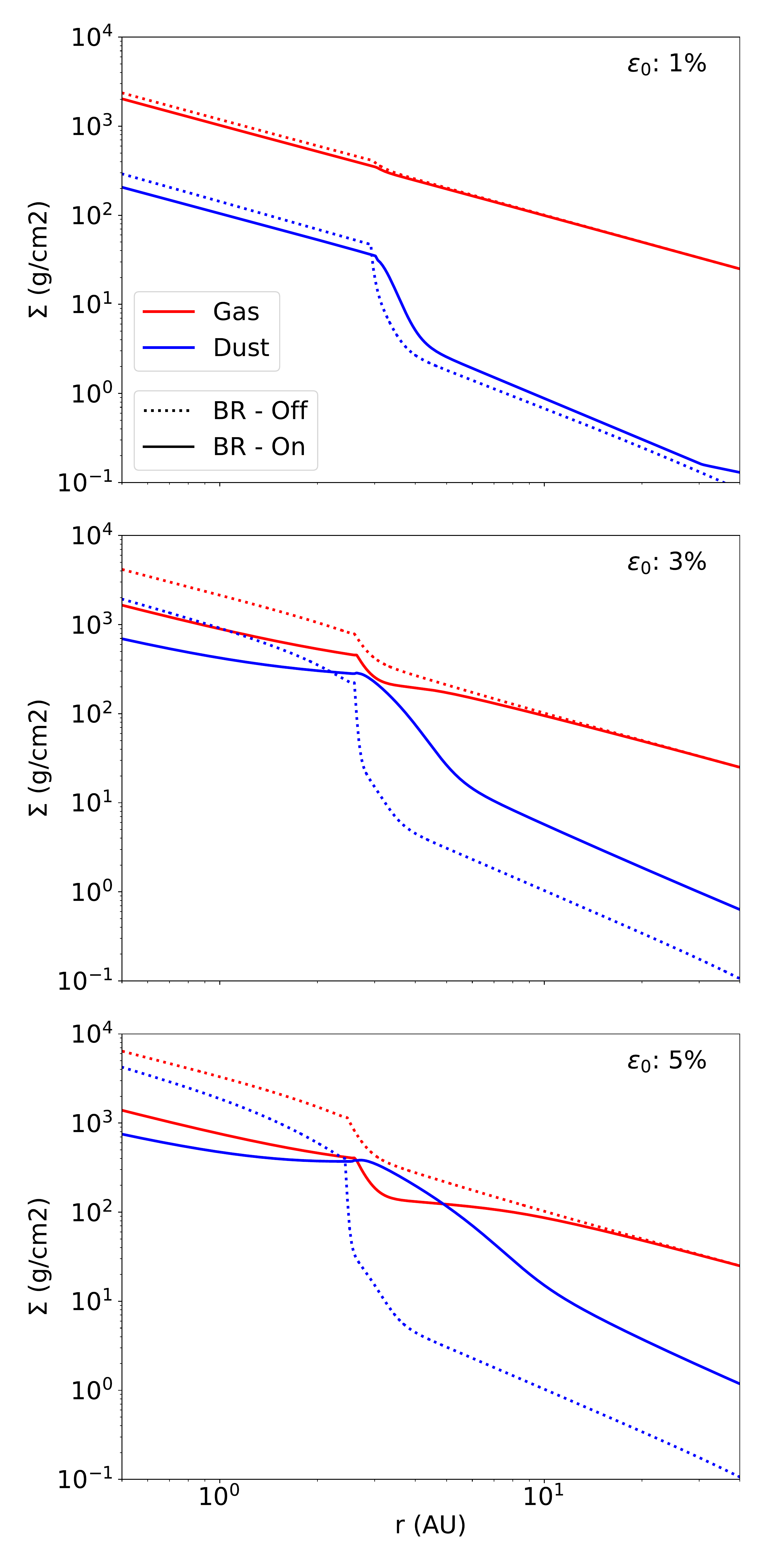}
 \caption{
 Comparison of the surface density profiles when the back-reaction is considered (solid lines) and ignored (dashed lines), after $\SI{0.4}{Myr}$.  For the cases with $\varepsilon_0 \geq 0.03$, the gas surface density is reduced when the back-reaction is considered in the inner regions, and the dust concentration is extended.
 }
 \label{Fig_BR_Off}
\end{figure}
\begin{figure}
\centering
\includegraphics[width=85mm]{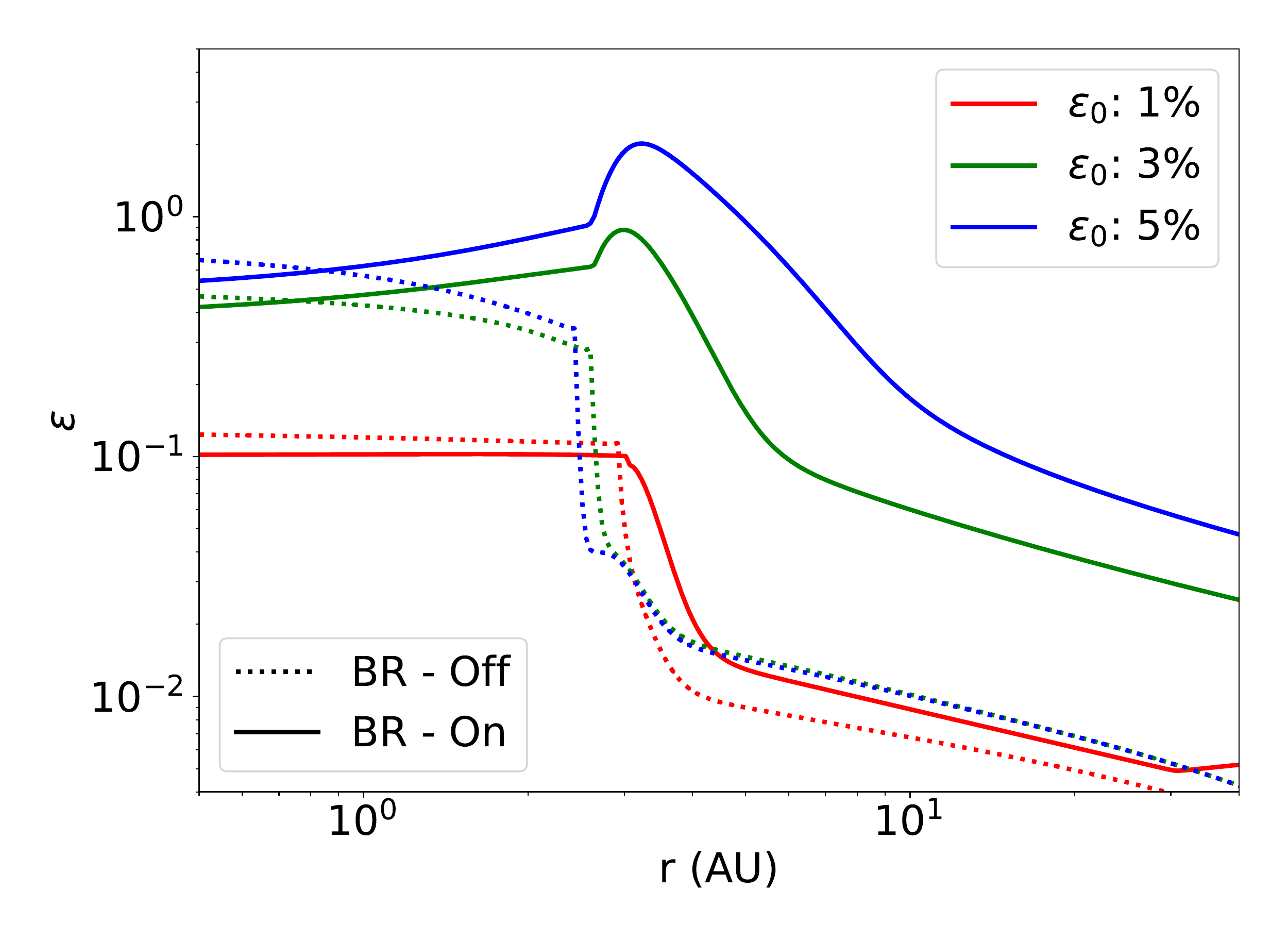}
 \caption{
 Comparison of the dust-to-gas ratio profiles when the back-reaction is considered (solid lines) and ignored (dashed lines), after $\SI{0.4}{Myr}$. When the back-reaction is ignored, the dust accumulates only inside the snowline.
 }
 \label{Fig_d2g_BR_Off}
\end{figure}
So far we have studied the impact on the dust back-reaction into the gas and dust density profiles, and in the gas velocity. So, how different is the situation when the back-reaction effect is ignored?\\
In \autoref{Fig_BR_Off} we turn off the back-reaction effects ($v_\textrm{g,r} = v_\nu$, $\Delta v_{\textrm{g},\theta} = - v_P$), and ignore the collective effect of dust on its diffusivity ($D_\textrm{d} = \nu$).
The simulation with $\varepsilon_0 = 0.01$ shows only minor differences, corresponding to a faster dust accretion. This is an indication that for low dust-to-gas ratios the back-reaction onto the gas is not important.\\
For the simulations with $\varepsilon_0 \geq 0.03$ we observe that, without the back-reaction effect, the dust only concentrates in the inner regions due to the traffic jam caused by the change in particle sizes at the snowline.
Accordingly, the water vapor delivered by the icy particles also increases the total gas content.\\ 
\autoref{Fig_d2g_BR_Off} shows how the dust-to-gas ratio profile is affected by the dust back-reaction. Only when the back-reaction is considered the solid particles can pile up outside the water snowline, due to the perturbed pressure gradient and the slower dust motion. 
For the simulations with $\varepsilon_0 \geq 0.03$, the dust back-reaction increases the dust-to-gas ratio by over an order of magnitude outside the water snowline.
This agrees with previous results of \cite{Drazkowska2017, Hyodo2019} where the dust back-reaction was incorporated as the collective drift of the dust species.
%
\subsection{The importance of the disk profile and size.}
\label{sec_Results_LBP}
\begin{figure} 
\centering
\includegraphics[width=85mm]{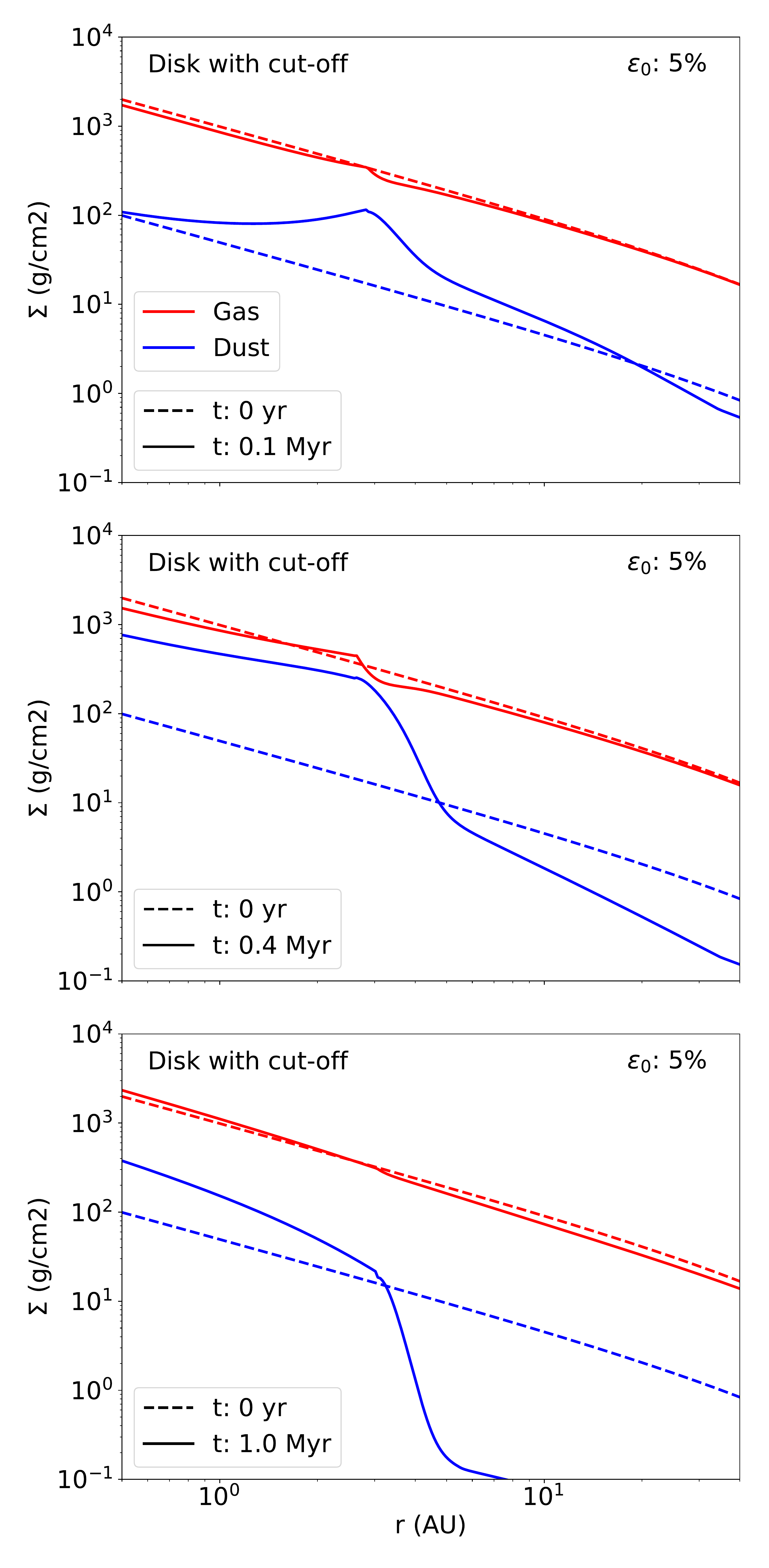}
 \caption{
 Surface density profiles of gas (red) and dust (blue) at different times (solid lines). The initial condition corresponds to the self-similar profile (dashed lines).
 \textit{Top:} The simulation initially behaves in the same way as the power law profile until $\SI{0.1}{Myrs}$.
 \textit{Mid:} At $\SI{0.4}{Myrs}$ the dust supply gets exhausted before the back-reaction push can further deplete the gaseous disk. 
 \textit{Bottom:} After \SI{1}{Myr}, the gas profile looks very similar to its initial condition, but most of the dust has been accreted.
 }
 \label{Fig_SurfaceEvolution_LBP}
\end{figure}
\begin{figure} 
\centering
\includegraphics[width=85mm]{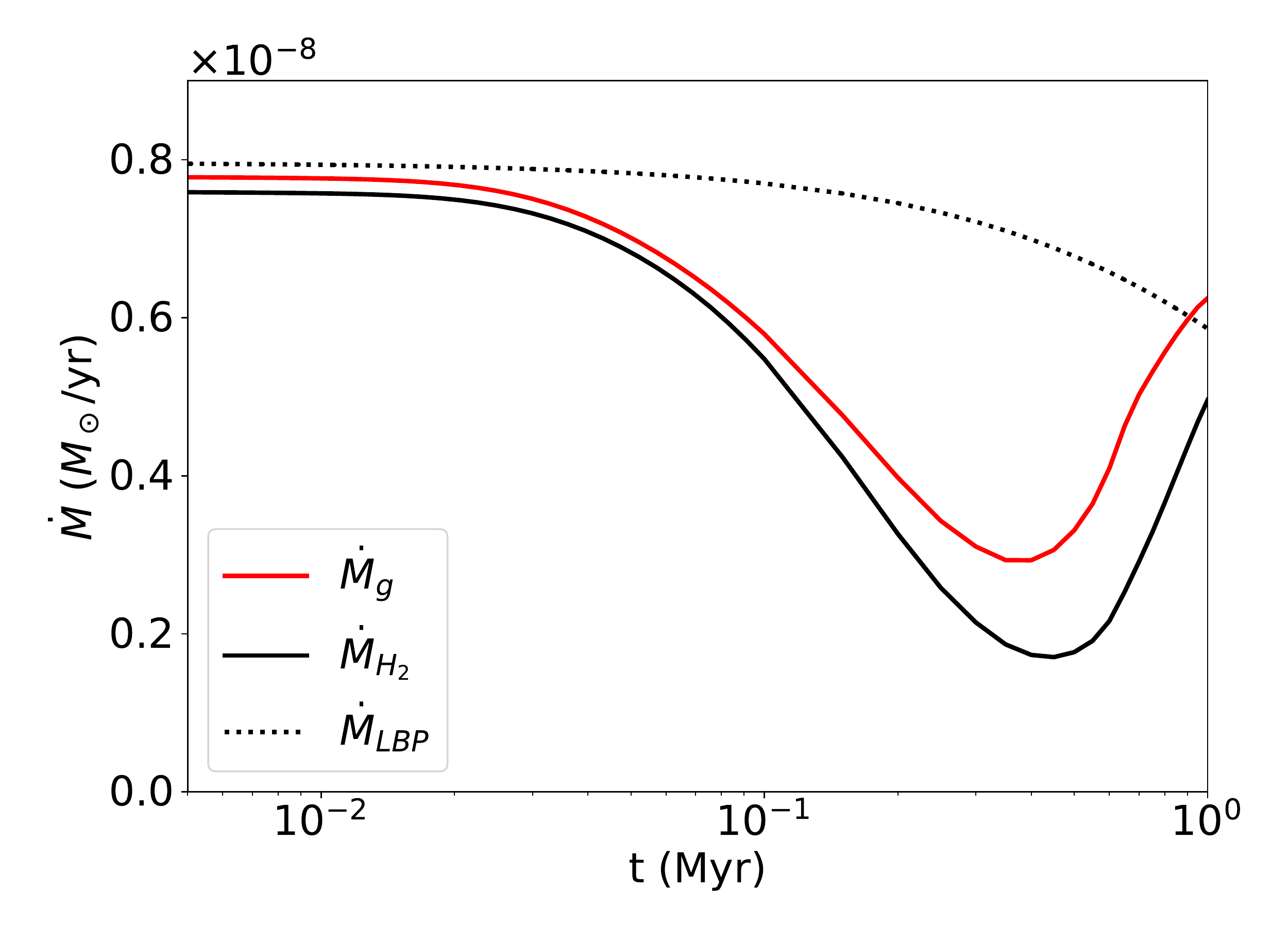}
 \caption{
 Accretion rate over time for the simulation with self-similar profile and $\varepsilon_0 = 0.05$. 
 The gas accretion rate (red) decreases as the dust back-reaction damps gas velocity, and rises again after the dust is depleted. The accretion rate of $\textrm{H}_2, \textrm{He}$ (black) is even lower, as the gas supply of the outer regions is reduced at the snowline. 
 The accretion rate of the standard self-similar solution (dotted line) is plotted for comparison.
 }
 \label{Fig_AccretionEvolution_LBP}
\end{figure}
How much the dust can perturb the gas surface density depends on the dust-to-gas ratio and the dust sizes, but also on how long the back-reaction is effectively acting.\\
In the \dbquote{High $\varepsilon_0$} case, the dust first creates a small depletion into the gas outside the snowline, the pressure slope changes and allows for large particles to further accumulate. Yet, this scenario assumes that icy particles are being constantly delivered towards the snowline, while in reality the supply has a limit given by the disk size.\\
We made a test simulation with $\varepsilon_0 = 0.05$ as in the \dbquote{High $\varepsilon_0$} case, but this time starting with a self-similar profile \citep{Lynden-Bell1974}, following:
\begin{equation} \label{eq_gas_LBPprofile}
    \Sigma_\textrm{g}(r) = \Sigma_0 \left(\frac{r}{r_0}\right)^{-p} \exp(-r/r_c),
\end{equation}
with a cut-off radius of $r_c= \SI{100}{au}$.\\ 
From \autoref{Fig_SurfaceEvolution_LBP} we can see the evolution of this simulation until $\SI{1}{Myr}$. 
Though we still observe that dust accumulates at the snowline, reaching dust-to-gas ratios between $\epsilon = 0.7 - 0.8$, and that the back-reaction push still creates a small dip in the gas surface density outside the snowline, the supply of solids is not enough to perturb the gas over extended periods of time. 
In this disk of limited size, no extended dust accumulation outside the snowline is observed.\\
The effect that still remains present is the decrease of the accretion rate (\autoref{Fig_AccretionEvolution_LBP}). As long as dust is delivered at the snowline, the accretion rate of gas is damped, and the mass fraction of the $\textrm{H}_2, \textrm{He}$ mixture is decreased in the inner regions.\\
We find that between $\SI{0.4}{} - \SI{0.5}{Myr}$ the dust concentration reaches its maximum value at the snowline (roughly the time required for the dust in the outer regions to grow and drift through the disk), and the accretion rate reaches its minimum  of $\SI{3.e-9}{M_\odot/yr}$, where only $60\%$ of the accretion flow corresponds to $\textrm{H}_2, \textrm{He}$.\\
After $\SI{1}{Myr}$ the dust is completely depleted, the disk surface density roughly recovers the self similar profile and the accretion rate rises back again.
%
%
\section{Discussion}\label{sec_Discussion}
%
\subsection{When is dust back-reaction important?}
So far we have seen that when the back-reaction is effective, it can enhance the dust concentration at the snowline (\autoref{Fig_d2gRatio}), damp the gas accretion rate (\autoref{Fig_AccretionDamp}), and deplete the inner regions from hydrogen and helium (\autoref{Fig_H2Abundance}).\\
All of these effects can be traced back to the push exerted by the dust back-reaction onto the gas (\autoref{eq_gas_vr}), that reduces the pressure gradient (which enhances dust accumulation), and slows down the flux of material from outside the snowline to the inner regions.\\
As a rule of thumb, the gas dynamic is altered whenever the pressure velocity term is comparable to the damped viscous velocity ($\bar{A} v_\nu \sim 2\bar{B} v_P$, \autoref{eq_gas_vr}), which occurs roughly when the particles have large Stokes number and high dust-to-gas ratios such that $\mathrm{St}\, \epsilon /(\epsilon + 1) \sim \alpha$ \citep{Kanagawa2017, Dipierro2018}.\\
In an inviscid disk ($\alpha_\nu \approx 0$), the gas velocity is dominated by the term $2\bar{B} v_P$, and the gas moves against the pressure gradient \citep{Tanaka2005}. 
On the other side, if the disk is highly turbulent ($\alpha_\nu \gg \mathrm{St \epsilon}$), then the gas evolves with a damped viscous velocity $\bar{A} v_\nu$. 
In \autoref{sec_Appendix_BackreactionCriteria} we include an equivalent criterion to determine the effect of the back-reaction, based on the angular momentum exchange between the dust and gas.\\
Through this paper we found that a high global dust-to-gas ratio of $\varepsilon_0 \gtrsim 0.03$, and a low viscous turbulence of $\alpha_\nu \lesssim \SI{e-3}{}$ (see \autoref{sec_Appendix_ParamSpaceExplore}), are necessary for the back-reaction push to perturb the combined evolution of gas and dust.\\
We also showed that the duration and magnitude of these effects depends on the disk size, as the dust accumulation and the perturbation onto the gas stop once the solid reservoir is exhausted (\autoref{Fig_SurfaceEvolution_LBP}). 
In particular, for a disk with cut-off radius of $r_c = \SI{100}{au}$ the dust drifts from the outer regions to the snowline in $\SI{0.4}{Myr}$. Afterwards, the back-reaction effects decay in a viscous timescale of the inner regions (roughly another $\SI{0.5}{Myr}$).\\
Moreover, part of the dust accumulated at the snowline will be converted into planetesimals through streaming instability \citep{Youdin2005, Drazkowska2017}, which in turn will reduce the dust-to-gas ratio and smear out the back-reaction effects.\\
We should keep in mind however, that the results presented in this paper only occur if the snowline acts as a traffic jam for dust accretion, which is caused by the difference in the fragmentation velocities of dry silicates and icy aggregates. Yet, recent studies suggest that there is no difference between the sticking properties of silicates and ices \citep{Gundlach2018, Musiolik2019, Steinpilz2019}, implying that the traffic jam should not form in the first place.
\subsection{Other scenarios where the back-reaction might be important}
Similar traffic jams and dust traps can occur in different regions of the protoplanetary disk. 
Given high dust concentrations and large particles sizes, the dust back-reaction may perturb the gas in locations such as dead-zones \citep{Kretke2009, Ueda2019, Garate2019}, the outer edge of gaps carved by planets \citep{Paardekooper2004, Rice2006, Weber2018}, and the edge of a photo-evaporative gap \citep{Alexander2007}.\\
In numerical models of protoplanetary disks, the back-reaction effects should be considered when estimating the gas accretion rate \citep[which is reduced by the interaction with the dust,][]{Kanagawa2017}, the planetesimal formation rate \citep[which would be enhanced for higher dust concentrations,][]{Drazkowska2017}, or the width of a dusty ring in the outer edge of a gap carved by a planet \citep{Kanagawa2018, Weber2018, Drazkowska2019}.\\
The effects of the back-reaction could actually become effective at later stages of the disk lifetime, provided that other mechanisms continue to trap the dust delivered from the outer regions, for example, if a planet forms from the planetesimal population at the water snowline \citep{Drazkowska2017}, it would carve a gap that can effectively trap dust particles \citep{Pinilla2012, Lambrechts2014}, and create a new environment where the back-reaction can affect the gas and dust dynamics \citep{Kanagawa2018}.\\
On smaller scales the dust back-reaction triggers the streaming instability, locally enhancing the concentration of dust particles until the solids become gravitationally unstable \citep{Youdin2005}, and close to the midplane the friction between layers of gas and dust results in a Kelvin-Helmholtz instability between the two components \citep{Johansen2006}.\\
Finally, one scenario that we did not cover in our parameter space is when the turbulence is so low ($\alpha_\nu = 0$) that the disk advection is reversed all the way to the inner boundary, which could lead to further perturbations at the snowline location, though a proper treatment of the dust sublimation should be included to account for this scenario.\\
Among our results, we could not reproduce the accumulation of dust in the outer regions of the disk described by \cite{Gonzalez2017}, as the dust particles drift towards the inner regions before creating any perturbation in the outer gas disk. We also find that by taking into account the growth limits, the back-reaction is less efficient than previously thought \citep{Kanagawa2017}, as the fragmentation barrier prevents the particles to grow to sizes beyond $\textrm{St}_\textrm{frag}$, and limiting the effect of the back-reaction even if the gas surface density decreases.\\
We do not expect our results to be significantly affected by changes in the disk mass or the stellar mass. 
Since particles sizes around the snowline are limited by the fragmentation barrier, the changes in any of these two parameters will only affect the physical size of the particles, but not their Stokes number (\autoref{eq_frag_limit}) which controls the dynamical contribution of the particles to the gas motion. The timescales and the snowline location would change accordingly, but the qualitative results presented in this work should hold true.
\subsection{Layered accretion by dust settling} \label{sec_LayeredAccretion}
\begin{figure} 
\centering
\includegraphics[width=100mm]{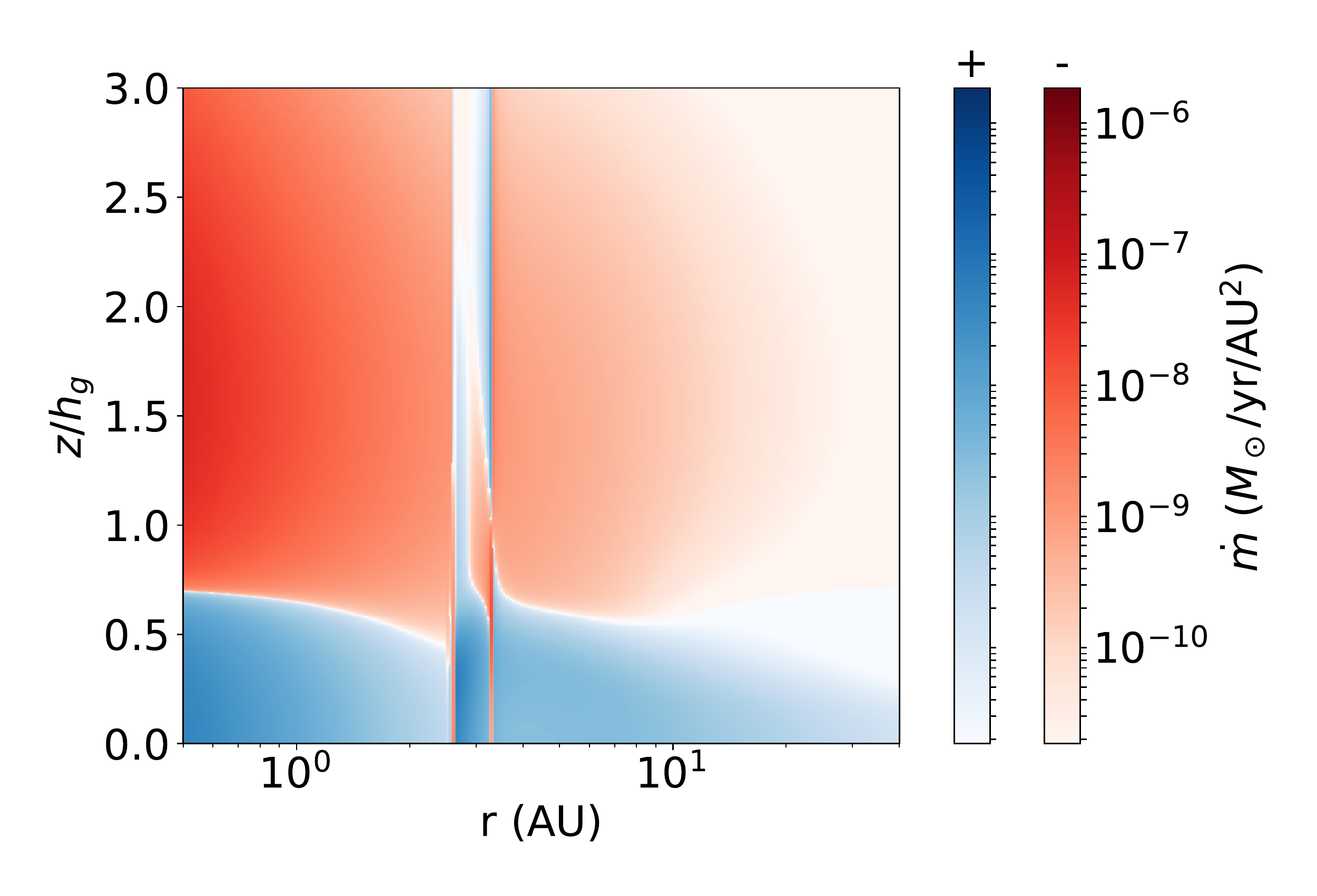}
\includegraphics[width=85mm]{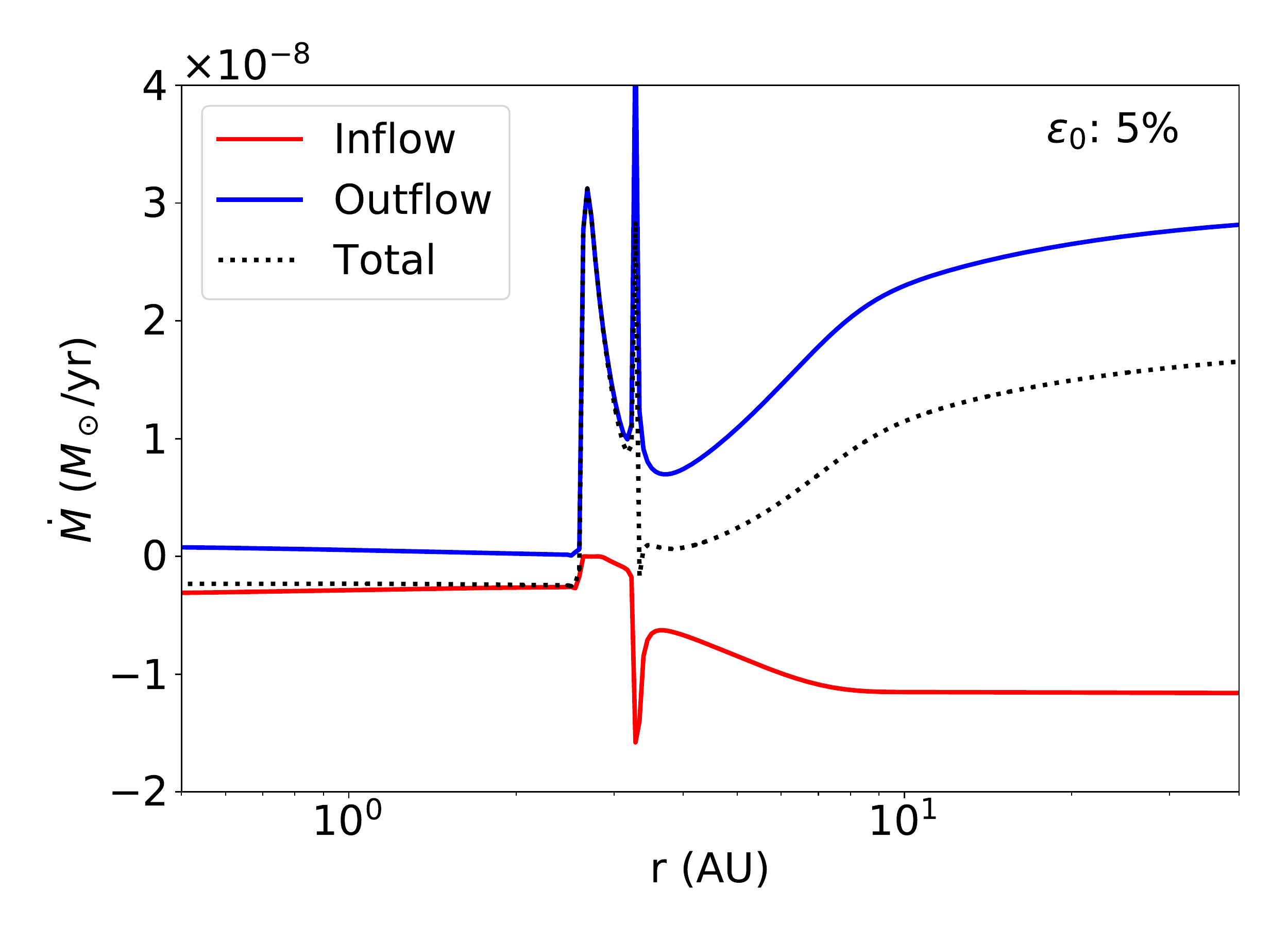}
 \caption{
 \textit{Top:} Mass flux for the simulation with $\varepsilon_0 = 5\%$, in the radial and vertical direction, obtained using the \cite{Takeuchi2002} vertical velocity profiles. The blue regions show the material outflow, and the red regions show the inflow.
\textit{Bottom:} Accretion inflow (red), outflow (blue), and the total mass accretion (dotted) profiles.}
 \label{Fig_LayeredAccretion}
\end{figure}
Because large particles settle towards the midplane, the back-reaction push onto the gas can be stronger at the disk midplane than at the surface \citep{Kanagawa2017}, which can result in the upper layers flowing inward (unperturbed by the dust), while the inner layers flow outward (due to the dust back-reaction). Depending on the particle sizes, this might result in different accretion rates at different heights.\\
While our approach to treat the vertical structure traces correctly the net mass transport (Section \ref{sec_VerticalApproximation}, \autoref{sec_Appendix_VerticalApproximation}) it does not provide information about layered accretion flow.
To check if there is a substantial inflow of material at the upper layers, we calculate the accretion rate at every height (using the vertical model from \cite{Takeuchi2002}, see \autoref{sec_Appendix_VerticalApproximation}) and measure the total mass inflow and outflow separately (\autoref{Fig_LayeredAccretion}).\\
We find that inside the snowline ($r < \SI{2.7}{au}$), where the dust particles are small ($\mathrm{St} \sim \SI{e-4}{}$) and well mixed with the gas, the back-reaction damps the gas motion uniformly at all heights, and the total inflow is of $\SI{3.e-9}{M_\odot/yr}$, only $\SI{6.e-10}{M_\odot/yr}$ higher than the net accretion rate onto the star.\\
In the regions beyond the snowline dust accumulation ($r > \SI{3.0}{au}$), we find that the accretion rate is layered, with the disk midplane flowing outward, while the surface layers move inward. The material inflow in this case is comparable to that of a dust free disk ($\dot{M} \sim \SI{e-8}{M_\odot/yr}$), even if the net mass flux is positive. This is in agreement with the results of \cite{Kanagawa2017}. 
For these regions, we find that the dust back-reaction can revert the gas flow up to $2 h_\textrm{d}$, which for the large dust particles at $\SI{3}{au}$ ($\mathrm{St} \sim \SI{e-2}{}$) corresponds to $0.6 h_\textrm{g}$.\\
Interestingly, at the snowline location where the dust particles accumulate ($\SI{2.7}{au} <r < \SI{3.0}{au}$), the dust back-reaction is strong enough to perturb the gas surface density.
The steeper negative surface density slope found at the snowline causes the viscous accretion to be reduced or reversed at all heights \citep{Takeuchi2002}. The accretion inflow is then reduced to a value of $\dot{M} = \SI{8.e-10}{M_\odot/yr}$ for the simulation with $\varepsilon_0 = 0.03$ (in which the reduced inflow only occurs above $0.7 h_\textrm{g}$), and to $\dot{M} = \SI{6e-12}{M_\odot/yr}$ for the simulation with $\varepsilon_0 = 0.05$, where the inflow only occurs $2.5 h_\textrm{g}$ above the midplane.\\
The steepening of the surface density slope at the water snowline was not observed in the previous results of \cite{Kanagawa2017} as they did not include the snowline or a dust growth and recondensation model. We find that this perturbation caused by the dust accumulation at snowline is key to reduce or stop the accretion inflow over a wide vertical range, which can be larger than the dust scale height itself.\\
Given that the disk mass inside the snowline is of $\SI{2.e-3}{M_\odot}$, the composition of the gas phase described in Section \ref{sec_depletionH2} should be corrected for the material flowing from the outer disk into the inner regions. 
For the simulation with $\varepsilon_0 = 0.03$ the $H_2 + He$ ratio should be higher by a $20\%$, considering that the inflow is reduced by over an order of magnitude at the snowline location (though not completely stopped). 
For the simulation with $\varepsilon_0 = 0.05$ all our results hold. 
\subsection{Observational Implications}
The perturbation caused by the dust back-reaction at the snowline is only effective if the viscous turbulence is low, if the dust-to-gas ratio is high, and only acts at early times of the disk evolution, while dust is supplied towards the inner regions. Given these constraints, we want to find which disk properties would fit in this parameter space, and what signatures we can expect to find if the back-reaction is effectively perturbing the gas.
\subsubsection{Ideal targets}
Young Class 0 and Class I disks seem to have typical sizes around \SI{100}{} - \SI{200}{au} \citep[][Table 1]{Najita2018}, so solids can be delivered to the inner regions only until $\SI{0.5}{} - \SI{1}{Myr}$, before the disk is depleted of dust (unless a pressure bump prevents particles from moving towards the star). This means that older disks ($t > \SI{1}{Myr}$) are unlikely to present any perturbation from the back-reaction push.\\
Then, among young disks and assuming viscous accretion, only those with low accretion rates of:
\begin{equation} \label{eq_low_accretion_rate_target}
    \Dot{M} \lesssim \SI{e-8}{M_\odot/yr} \left(\frac{M_\textrm{disk}}{\SI{0.1}{M_\odot}}\right) \left(\frac{r_c}{\SI{100}{au}}\right)^{-1} \left(\frac{M_\textrm{star}}{\SI{}{M_\odot}}\right)^{-1/2} \left(\frac{T_0}{\SI{300}{K}}\right),
\end{equation}
could be subject to the back-reaction damping, as a low viscous evolution ($\alpha_\nu \lesssim \SI{e-3}{}$) is required for the dust to affect the gas. 
In terms of the dimensionless accretion parameter introduced by \cite{Rosotti2017}, defined as:
\begin{equation} \label{eq_Rosotti_eta_param}
    \eta = \frac{\tau \Dot{M}}{M_\textrm{disk}},
\end{equation}{}
a disk of age $\tau$ would require $\eta \lesssim 0.1$ for the dust back-reaction to effectively perturb the gas. 
\subsubsection{On the gas orbital velocity}
If the concentration of dust in any region is high, then the gas pressure support is reduced and the orbital velocity approaches to the keplerian velocity $v_K$ (\autoref{eq_gas_vt}).\\
At the midplane, where large grains concentrate, the gas motion deviates from the keplerian velocity by:
\begin{equation} \label{eq_gas_vt_MidplaneApprox}
    \Delta v_{\textrm{g},\theta} \approx - \frac{v_P}{1 + \max(1, \sqrt{\mathrm{St}/\alpha_t})\cdot \epsilon } ,
\end{equation}
where the $\sqrt{\mathrm{St}/\alpha_t}$ factor measures the concentration of large particles at the midplane by settling (see \autoref{sec_Appendix_VerticalApproximation}). 
If in our disk the initial pressure velocity around the snowline was $v_P \approx \SI{2e-3}{} v_K$, then the dust back-reaction and the accumulation of water vapor makes the gas orbit at velocities of  $\Delta v_{\textrm{g},\theta} \approx \SI{7e-4}{} v_K$.
At $\SI{2.7}{au}$, where the snowline is located in our simulations, this correspond to a difference from the keplerian velocity of approximately $\Delta v_{\textrm{g},\theta} \approx \SI{10}{m/s}$.\\
We expect that in future observations, the deviations from the keplerian velocity could be used to constrain the dust content.
\cite{Teague2018} already showed that the deviations from the keplerian velocity can be used to kinematically detect a planet, reaching a precision of $\SI{2}{m/s}$. Better characterizations of the orbital velocity profiles in dust rings may then be used to differentiate between a planet perturbation and a dust back-reaction perturbation, based on the profile shape.\\
Unfortunately, the spatial resolution required to observe this variation is less than \SI{10}{mas}, for a disk at a distance of \SI{100}{au} and a snowline at \SI{3}{au} from the star, and next generation instruments would be required. The velocity deviation could be easier to detect for disks around Herbig stars where the snowline is located at larger radii.
\subsubsection{Shadows casted by dust accumulation}
A recent study of \cite{Ueda2019} showed that dust can accumulate at the inner edge of a dead zone \citep[a region with low ionization and low turbulence, ][]{Gammie1996}, and cast shadows that extend up to $\SI{10}{au}$. \\
We notice that our accumulation of dust at the snowline is similar to the dead zone scenario, in the sense that high dust-to-gas ratios are reached in a narrow region of the inner disk (\autoref{Fig_d2gRatio}). 
Therefore, we hypothesize that similar shadows could be found in the regions just outside the snowline if enough dust is present. 
Still, radiative transfer simulations would be needed to determine the minimum dust-to-gas ratio necessary to cast a shadow.
\subsubsection{Effects of the snowline traffic jam}
The fast drift of the icy particles particles and the traffic jam at the snowline results in the accumulation of both small silicate dust and water vapor inside the snowline, even if the effect of the dust back-reaction is ignored (see \autoref{Fig_BR_Off} and \ref{Fig_d2g_BR_Off}).\\
We find that if the initial dust supply is large enough (high $\epsilon_0$ and large disk size), then during the early stages of the disk evolution ($t \lesssim \SI{1}{Myr}$) we can expect the material accreted into the star to be rich in silicates and refractory materials carried by the dust (see \autoref{Fig_d2gRatio}), rich in oxygen (which is carried by the water vapor), and relatively poor in hydrogen, helium, and other volatile elements mixed with the gas outside the water snowline (see \autoref{Fig_H2Abundance}), such as nitrogen and neon.
The X-ray emission could provide estimates of the abundance ratios in the accreted material \citep{Gunther2006}, though the coronal emission of neon in young stars could mask some of these abundances (H. M. G\"unther, private communication).\\
The increased concentration of water vapor in the warm inner regions would also enhance the emission from the water rotational lines. These lines have been already detected in different disks \citep{Carr2008, Salyk2008} in the mid-IR with Spitzer IRS, and could be further observed in the future using Mid-Infrared Instrument at the James Webb Space Telescope \citep[MIRI,][]{Rieke2015}.\\
Additionally, the excess of water should lead to low C/O ratios inside the snowline for young protoplanetary disks \citep{Oberg2011, Booth2018}.
%
\section{Summary}\label{sec_Summary}
In this study we included the effects of the dust back-reaction on the gas in a model of the water snowline, which is known to act as a concentration point for dust particles due to the change in the fragmentation velocity between silicates and ices, and the recondensation of water vapor into the surface of icy particles \citep{Drazkowska2017}.\\
Our model shows how the dust back-reaction can perturb the gas dynamics and disk evolution, though the parameter space required for this to happen is limited.\\
In the vicinity of the snowline, provided that the global dust-to-gas ratio is high ($\varepsilon_0 \gtrsim 0.03$) and the viscosity low ($\alpha_\nu \lesssim \SI{e-3}{}$), the effects of the dust back-reaction are:
\begin{itemize}
    \item Revert the net gas flux outside the snowline. 
    \item Reduce the gas inflow at the snowline by over an order of magnitude.
    \item Damp the gas accretion rate onto the star to a $30\% - 50\%$ of its initial value.
    \item Reduce the hydrogen-helium content in the inner regions, and concentrate water vapor at the snowline.
    \item Concentrate solids at the snowline reaching dust-to-gas ratios of $\epsilon \gtrsim 0.8$.
\end{itemize}
These effects build up as long as dust is supplied from the outer disk into the snowline, with the duration set by the growth and drift timescale of the outer regions. After the dust reservoir is exhausted, the back-reaction effects decay in the viscous timescale of the inner regions. For a disk with size $r_c = \SI{100}{au}$, we find that dust accumulates only during the first $\SI{0.4}{Myr}$, and that the perturbation onto the gas has disappeared by the age of $\SI{1}{Myr}$.\\
The high dust-to-gas ratios required to trigger the back-reaction effects, and the traffic jam at the snowline, can result in an enhanced water content in the inner regions, in the accretion onto the star to be enriched with refractory materials and oxygen, and perhaps a shadow to be casted outside the snowline location by the accumulation of dust particles.\\
Other types of dust traps could present similar behaviors, though each case must be revisited individually to evaluate the magnitude of the perturbation of the back-reaction into the gas velocity.
\begin{acknowledgements}
We would like to thank S. Facchini, H. M. G\"unther, and G. Rosotti for their comments and discussions during the early stages of this manuscript. We also want to thank the anonymous referee for his/her comments, that improved the extent and clarity of this work.
The authors acknowledge funding from the European Research Council (ERC) under the European Union’s Horizon 2020 research and innovation programme under grant agreement No 714769, by the Deutsche Forschungsgemeinschaft (DFG, German Research Foundation) Ref no. FOR 2634/1, and by the Deutsche Forschungsgemeinschaft (DFG, German Research Foundation) under Germany's Excellence Strategy – EXC-2094 – 390783311.
\end{acknowledgements}
\bibpunct{(}{)}{;}{a}{}{,} 
\bibliographystyle{aa} 
\bibliography{TheBibliography.bib} 
%
%
%
\begin{appendix}
\section{Semi-analytical test for back-reaction simulations.} \label{Sec_Appendix_EquivalentTest}
In this section we intend to rewrite the radial velocity of the gas (\autoref{eq_gas_vr}) in a similar way to the standard viscous velocity of \cite{Lynden-Bell1974}.
The viscous velocity and the pressure velocity (\autoref{eq_visc_velocity} and \ref{eq_press_velocity}) can be rewritten in the following form:
\begin{equation} \label{eq_viscous_velocity_alter}
    v_\nu = -3 \alpha_\nu\frac{c_s^2}{v_K}  \gamma_\nu,
\end{equation}
\begin{equation} \label{eq_pressure_velocity_alter}
    v_P = -\frac{1}{2} \frac{c_s^2}{v_K}  \gamma_P,
\end{equation}
with $\gamma_\nu = \partiallogdiff{(\nu \,     \Sigma_\textrm{g} \, \sqrt{r})}$ and $\gamma_P = \partiallogdiff{P}$.\\
Using these expressions, we can rewrite the gas radial velocity (\autoref{eq_gas_vr}) as the viscous velocity in \autoref{eq_viscous_velocity_alter}, with the following $\alpha_\nu$-equivalent parameter:
\begin{equation} \label{eq_alpha_equivalent}
    \alpha_\textrm{eq} = A \alpha_\nu+ \frac{\gamma_P}{3 \gamma_\nu} B,
\end{equation}
\begin{equation} \label{eq_gas_velocity_alter}
    v_\textrm{g} = -3 \alpha_\textrm{eq}\frac{c_s^2}{v_K}  \gamma_\nu.
\end{equation}
This means that we can understand the evolution of a gas disk considering dust back-reaction, as a viscous evolution with a modified $\alpha_\nu$ value (we discuss the limits of this interpretation in Appendix \ref{Appendix_AlphaEq_Limits}). 
From this point we can make further simplifications to develop a semi-analytical test for a back-reaction simulation using a standard viscous evolution model.\\
Our first simplification is that the surface density and temperature follow a power law profile with $\Sigma \propto r^{-p}$ and $T \propto r^{-q}$, which sets the factor $\gamma_P/(3 \gamma_\nu)$ involving the density and temperature gradients to:
\begin{equation} \label{eq_gamma_powerlaw}
    \frac{\gamma_P}{3 \gamma_\nu} = - \frac{2 p + q + 3}{6 (2 - p - q)}.
\end{equation}
In particular, if the disk is in steady state with $p = 1$ and $q = 1/2$, then $\gamma_P/(3 \gamma_\nu) = -11/6$, and the accretion rate is:
\begin{equation} \label{eq_steady_accrate}
    \Dot{M} = 3\pi \alpha_\textrm{eq} \frac{c_s^2}{\Omega_K}\Sigma_g.
\end{equation}
Now, assuming that the distribution of dust particles has sizes between $0 < \mathrm{St} < \mathrm{St}_\textrm{max}$, and that $\St^2 \ll \epsilon$, we can constrain the value of $\alpha_\textrm{eq}$ using the single particle approximation for the coefficients $A_\textrm{single}$ and $B_\textrm{single}$ (\autoref{eq_backreaction_A_single} and \ref{eq_backreaction_B_single}). 
Then the minimum value that $\alpha_\textrm{eq}$ can take, given by the largest size particles, is: 
\begin{equation} \label{eq_alpha_equivalent_approx}
    \alpha_\textrm{eq, min} \approx  \frac{\alpha_\nu}{\epsilon +1} - \frac{11}{6} \frac{\epsilon\, \mathrm{St}_\textrm{max}}{(\epsilon +1)^2},
\end{equation}
and the maximum value that $\alpha_\textrm{eq}$ can take, given by the smallest particles with $\mathrm{St}\approx 0$, is:
\begin{equation} \label{eq_alpha_equivalent_approx_small}
    \alpha_\textrm{eq, max} \approx  \frac{\alpha_\nu}{\epsilon +1}.
\end{equation}
%
%
\subsection{Setting up a test simulation}
\begin{figure*}
\centering
\includegraphics[width=170mm]{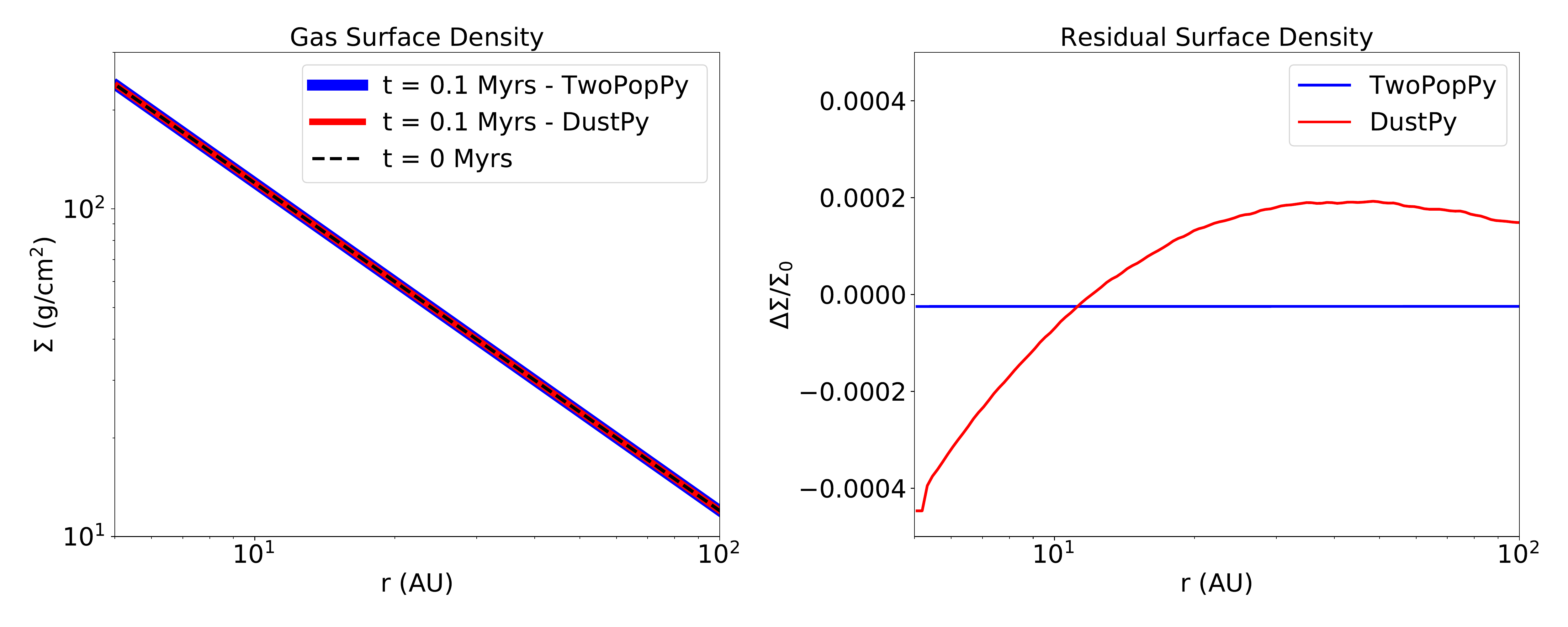}
 \caption{
 \textit{Left:} Initial and final surface density of the test simulations. If the disk evolution with back-reaction is equivalent to a regular viscous evolution, the steady state should be maintained through the simulation. 
 \textit{Right:} Surface density residuals relative to the initial state. After $\SI{0.1}{Myrs}$ of evolution, the simulations deviate by less a value of $0.05 \%$ from the steady state profile.
 }
 \label{Fig_Appendix_TestSurfaceDensity}
\end{figure*}
\begin{figure*}
\centering
\includegraphics[width=170mm]{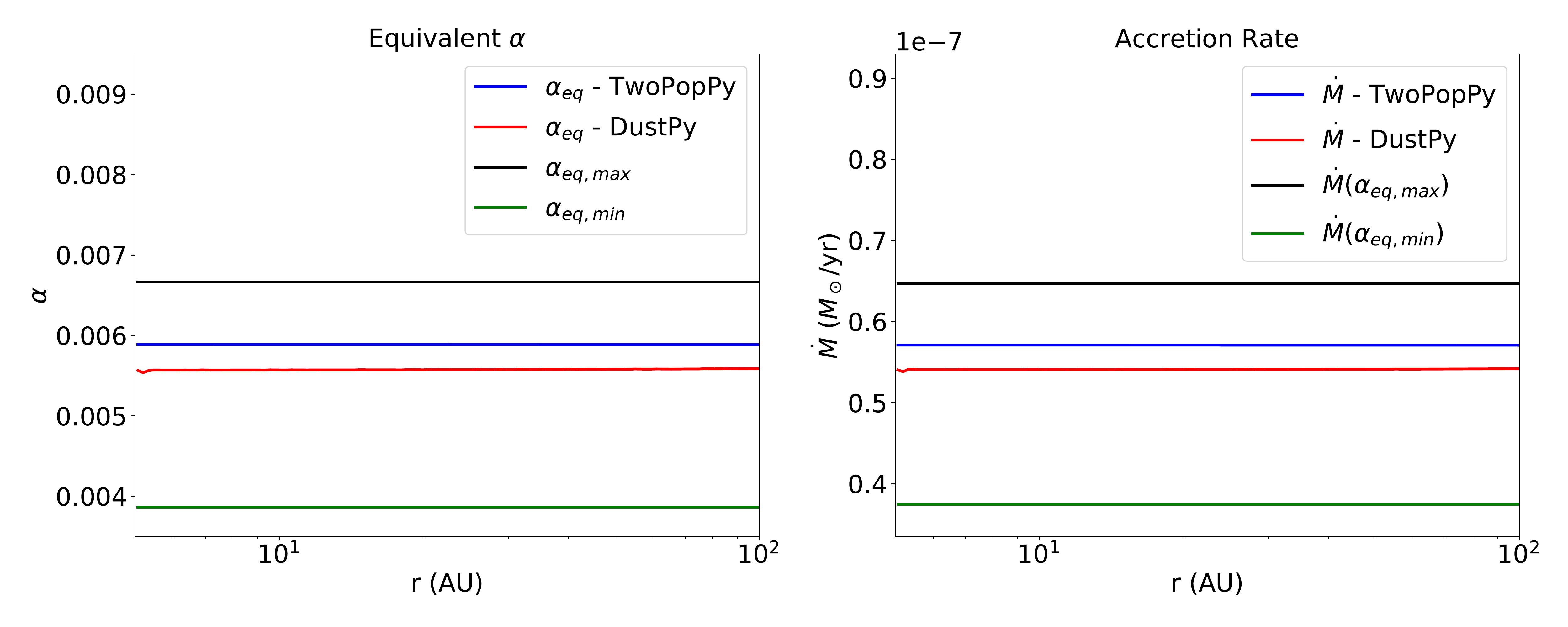}
 \caption{
 \textit{Left:} Equivalent $\alpha_\nu$ value (\autoref{eq_alpha_equivalent}) obtained from the simulations (red - DustPy, blue - TwoPopPy), and the analytical limits given by $\alpha_\textrm{eq, min}$ (green) and $\alpha_\textrm{eq, max}$ (black). 
 The value obtained from the simulations is in between the two limits, in agreement with the analytical model.
 \textit{Right:} Accretion rate measured from the simulations, and the steady state accretion rate for the different $\alpha_\textrm{eq}$ limits. 
}
 \label{Fig_Appendix_TestAccretionRate}
\end{figure*}
From the equivalent viscosity equation (\autoref{eq_alpha_equivalent}) we can set a test to ensure that the back-reaction effects in a numerical simulation are acting according to the theoretical model.\\
We prepare a test for the code \texttt{twopoppy} that was used throughout the paper \citep{Birnstiel2012}, and also for the code \texttt{DustPy} (Stammler and Birnstiel, in prep.), that solves the Smoluchowski equation for particle growth by sticking and fragmentation of multiple dust species as in \cite{Birnstiel2010}, along with the advection-diffusion equations (\autoref{eq_gas_advection} and \ref{eq_dust_advection}).\\
The test disk has the following set-up:
\begin{itemize}
    \item The surface density and temperature have steady state power law profiles with $p = 1$ and $q = 1/2$.
    \item To enhance the back-reaction damping and obtain obvious deviations from the regular dust-free evolution we set an unrealistic disk with $\epsilon = 0.5$.
    \item The fragmentation velocity follows $v_\textrm{frag} \propto r^{-q}$ so that the maximum particle size (\autoref{eq_frag_limit}) has a constant value of $\mathrm{St}_\textrm{max} = \SI{5e-3}{}$.
    \item The viscous turbulence is set to $\alpha_\nu = \SI{e-2}{}$, so that the back-reaction is not strong enough to reverse the accretion of gas.
    \item The dust diffusion is turned off, so that the dust is only advected through the velocity $v_\textrm{d,r}$ (\autoref{eq_dust_vr}).
    \item The disk is initialized with a fully grown particle distribution (so that the back-reaction effects are uniform through the disk).
    \item The back-reaction coefficients (in this test case) are implemented assuming that the dust-to-gas ratio is vertically uniform.
\end{itemize}
If the simulations are working properly, then the disk will remain in steady state, and the accretion rate will be constant in radius with a value given by the damped equivalent viscosity $\alpha_\textrm{eq}$ (\autoref{eq_steady_accrate}).
Since in this test case all the particles are small ($\St < \alpha_\nu$) and the size distribution is constant with radius, the dust-to-gas ratio and the back-reaction effects should also remain approximately uniform in time.\\
As shown in \autoref{Fig_Appendix_TestSurfaceDensity}, after $\SI{0.1}{Myr}$ the disk surface density between $\SI{5}{}-\SI{100}{au}$ remains close to the steady state profile, with a deviation of less than $0.1\%$ relative to its initial value.\\
\autoref{Fig_Appendix_TestAccretionRate} shows that the mass accretion rate of the gas in the simulations is $\Dot{M} \approx \SI{5.6e-8}{M_\odot/yr}$ and constant through the disk, in agreement with a steady state solution. More importantly, the value of the accretion is constrained between the minimum and maximum values given by $\alpha_\textrm{eq, min}$ and $\alpha_\textrm{eq, max}$ and \autoref{eq_steady_accrate}.\\
In terms of the viscous accretion, the back-reaction effect in our setup is equivalent to reduce the viscous turbulence $\alpha_\nu$ to a value of $\alpha_\textrm{eq} \approx 0.57\, \alpha_\nu$.\\
Both \texttt{twopoppy} and \texttt{DustPy} deliver similar results, with a relative difference of roughly $5 \%$ in the $\alpha_\textrm{eq}$ and $\Dot{M}$ values. 
From here we can conclude that the back-reaction effects observed in the two population model are expected to be in agreement with those from a proper particle distribution.
\subsection{Where the viscous approximation breaks} \label{Appendix_AlphaEq_Limits}
While we can always write the gas velocity in the form of \autoref{eq_gas_velocity_alter} using the $\alpha_\textrm{eq}$ parameter (\autoref{eq_alpha_equivalent}), the global disk evolution will still differ from a regular viscous evolution (unless $\alpha_\textrm{eq} \propto \alpha_\nu$), as the value of $\gamma_\nu$ does not depend on the slope of $\alpha_\textrm{eq}$.\\
In particular, the back-reaction effects cannot be treated as a viscous process if $\mathrm{St}\, \epsilon /(\epsilon + 1) \gtrsim \alpha_\nu$ \citep{Dipierro2018}. In this case the back-reaction push becomes more important than the inward viscous transport, and results in negative equivalent $\alpha_\textrm{eq}$ values, meaning that mass will be transported against the pressure gradient.\\
Also, in the outer regions of the disk where the surface density profile becomes steeper \citep[as in the self-similar solution][]{Lynden-Bell1974}, the viscous evolution spreads the gas outwards ($\gamma_\nu < 0$, $v_\nu > 0$). In these regions the dust back-reaction pushes the gas in the same direction as the viscous spreading ($2B v_P > 0$), and therefore contributes to evolve the outer disk faster than the inner disk.
\section{Modeling the vertical structure} \label{sec_Appendix_VerticalApproximation}
In section \ref{sec_VerticalApproximation} we discussed how to obtain the gas radial velocity from the net mass flux. For our simulations we considered the effect of the vertical structure of the gas and the settling of the dust on the back-reaction coefficients, but ignored the vertical profile of the pressure velocity $v_P(z)$ and viscous velocity $v_\nu(z)$. 
In this appendix we show that our results hold if we assume a standard vertical profile for the viscous and pressure velocity, and why our simple approximation works in the same way.
\subsection{Vertical profiles for $v_\nu$ and $v_P$}
\begin{figure} 
\centering
\includegraphics[width=85mm]{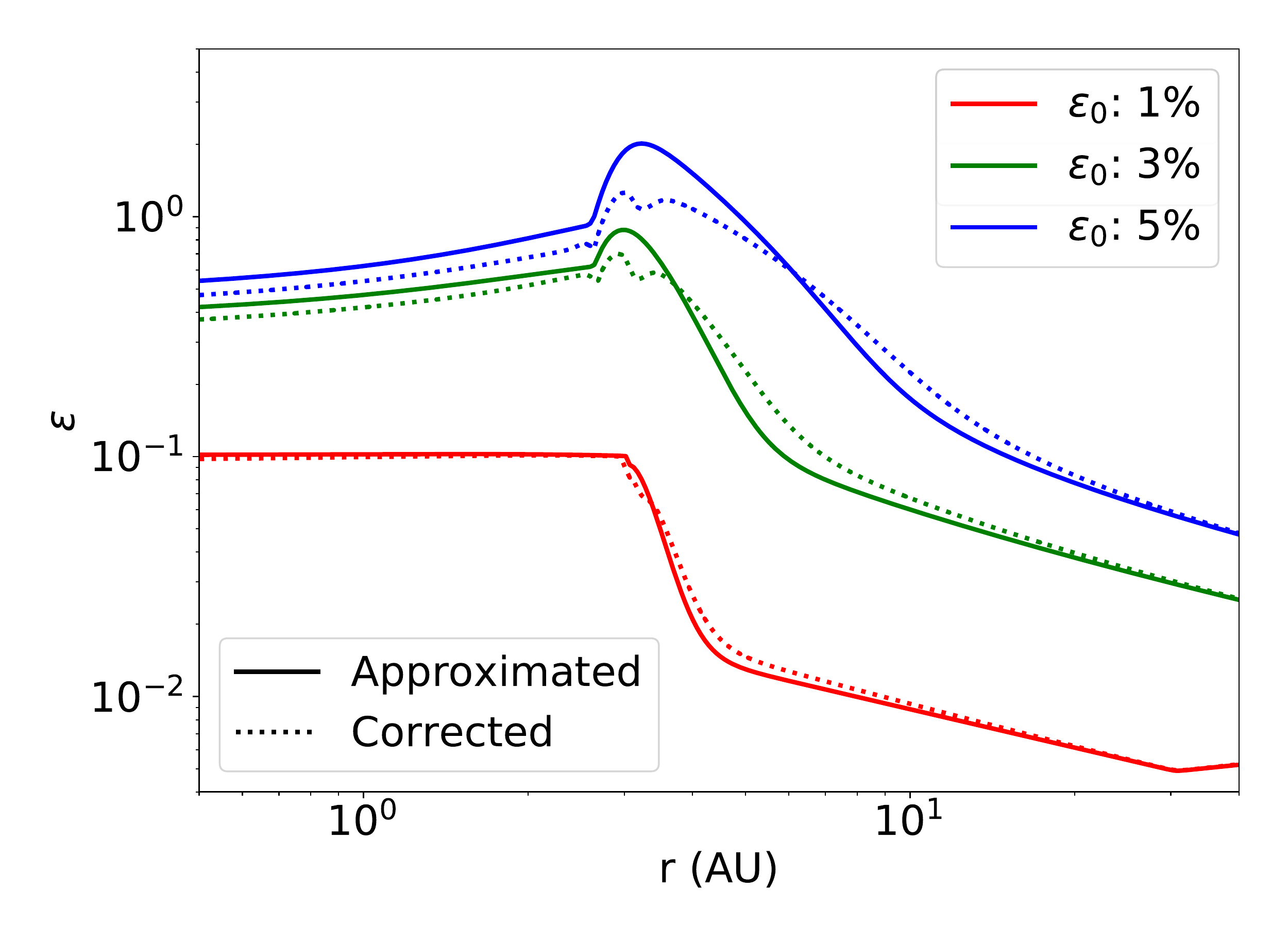}
\includegraphics[width=90mm]{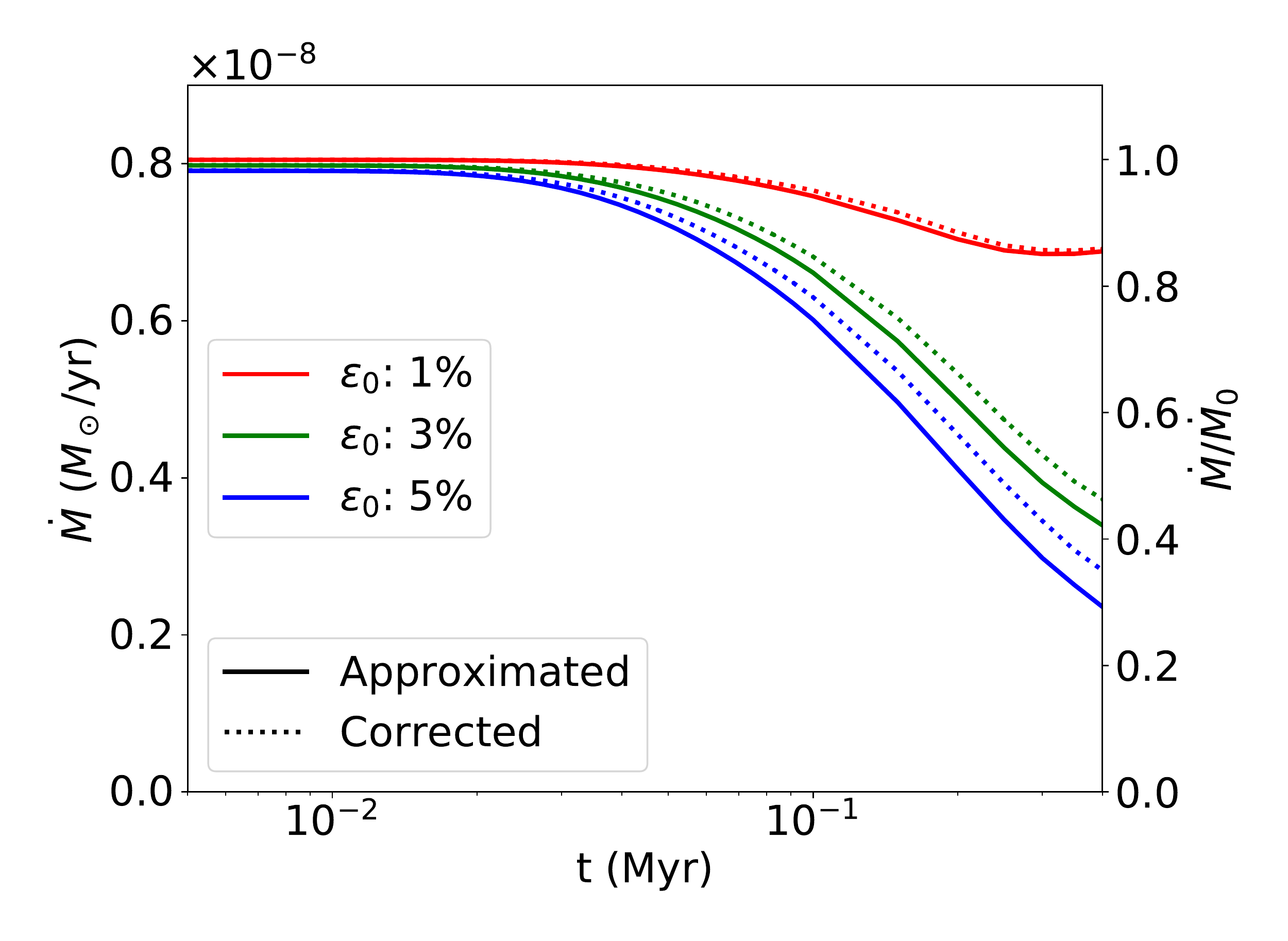}
 \caption{
 Comparison of the main results (\textit{top:} dust-to-gas ratio profiles, \textit{bottom:} accretion rate evolution) using our simple approximation for the vertical structure (solid lines) vs. the \cite{Takeuchi2002} vertical structure model (dotted lines).
 }
 \label{Fig_RefereeComparison_results}
\end{figure}

\begin{figure} 
\centering
\includegraphics[width=85mm]{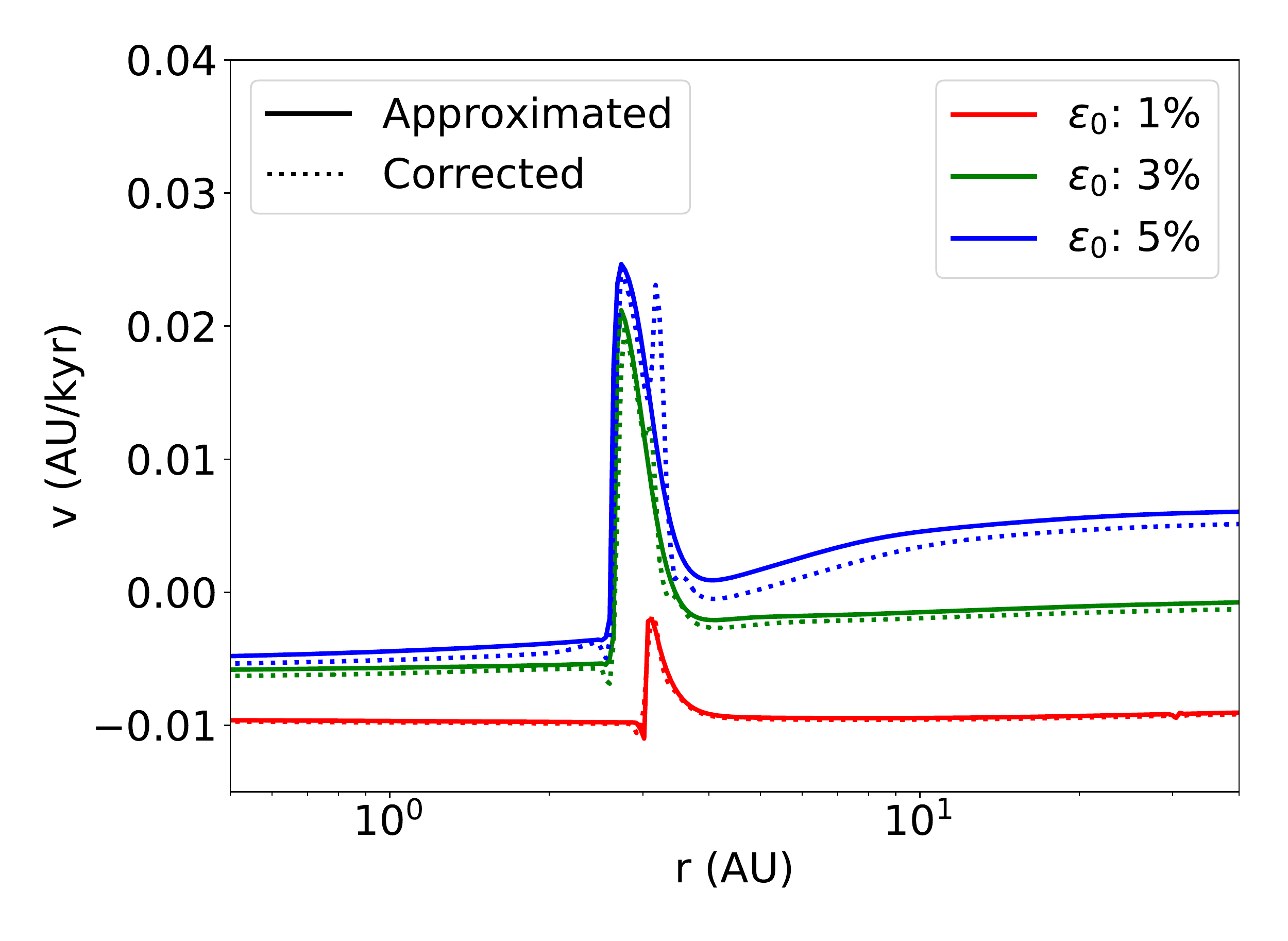}
 \caption{Gas radial velocity profiles (derived from the net mass flux, \autoref{eq_mass_flux_velocity}) using our simple vertical structure approximation vs. \cite{Takeuchi2002} model (\autoref{eq_viscous_velocity_takuchi} and \ref{eq_pressure_velocity_takuchi}). 
 }
 \label{Fig_RefereeComparison_velocity}
\end{figure}

Following the \cite{Takeuchi2002} model \citep[see also ][]{Kanagawa2017, Dipierro2018}, the vertical velocity profiles of $v_\nu$, $v_P$ are:
\begin{equation} \label{eq_viscous_velocity_takuchi}
    v_\nu(z)= \frac{\nu}{2 r} \left(6p + q - 3 + (5q - 9) \left(\frac{z}{h_\textrm{g}}\right)^2 \right),
\end{equation}
\begin{equation} \label{eq_pressure_velocity_takuchi}
    v_P(z)= v_K \left(\frac{h_\textrm{g}}{r}\right)^2 \left(p + \frac{q + 3}{2} + \frac{q - 3}{2}\left(\frac{z}{h_\textrm{g}}\right)^2 \right),
\end{equation}
where in this case $p$ and $q$ are the local exponents of the gas surface density and temperature profiles.\\
We include this information in the gas and dust velocity derived from the mass fluxes (\autoref{eq_mass_flux_velocity} and \ref{eq_mass_flux_velocity_dust}) and check how our results are affected.\\
\autoref{Fig_RefereeComparison_results} shows the dust-to-gas ratio profile, and the evolution of the gas accretion present only minor differences when considering the vertical structure for the viscous and pressure velocities (\autoref{eq_viscous_velocity_takuchi} and \ref{eq_pressure_velocity_takuchi}). 
For the simulations with $\varepsilon \geq 0.03$, only at the snowline location the dust-to-gas ratio is more spread over radii, reducing the maximum concentration by a factor of a few. Consequently, the accretion rate onto the star is approximately a $5\%$ higher when using the \cite{Takeuchi2002} prescription.\\
We find that our approximation reproduces well the radial velocity profile calculated with the \cite{Takeuchi2002} model (\autoref{Fig_RefereeComparison_velocity}), except in a narrow region beyond the snowline where the change in the gas surface density slope creates a spike in the gas velocity. 
However, this variation does not alter the rest of the simulation fields, and our results are maintained independently of the vertical structure prescription used for the viscous and pressure velocities.
\subsection{Explaining the vertical approximation}
There are two distinguishable regimes for the for gas and dust interaction.\\
The first regime is the one in which the particles are small ($\mathrm{St} < \alpha_\nu$). In this case the particles are well mixed with the gas, and therefore the dust-to-gas ratio (and the back-reaction coefficients) is uniform in the vertical direction. 
In this regime the back-reaction pushing term is also negligible, and the gas velocity is well approximated by the damped viscous velocity $\bar{v}_{\textrm{g},r} \approx \bar{A} v_\nu$.\\
The second regime is the case in which the particles are large ($\mathrm{St} \gtrsim \alpha_\nu$). In this case the dust settles towards the midplane and the back-reaction push becomes important. 
The dust particles near the midplane will move the gas against the local pressure gradient with a velocity of $v_{\textrm{g}, r} (z \lesssim h_\textrm{d}) \approx 2 \bar{B} (z \lesssim h_\textrm{d}) v_P$. Meanwhile, the upper layers of the disk have a low dust concentration, and allow the gas to move inward with the viscous velocity, therefore the gas velocity above the characteristic dust scale height can be approximated by $v_{\textrm{g}, r} (z \gtrsim h_\textrm{d}) \approx v_\nu$.
Then, the velocity derived from the net flux (\autoref{eq_mass_flux_velocity}) will yield a good approximation considering the upper layers flowing inward (dominated by the viscous flow) and the midplane layers flowing against the pressure gradient (dominated by the back-reaction push).\\
This approximation seems to be valid for most of the disk, except on a narrow region where slope of the gas density is reversed (at $r \approx \SI{3}{au}$), however this seem to be more related to a resolution problem than a physical reason, the surface density slope was smoothed in this case to avoid numerical problems in this region. Because of this region we prefer using an approximate solution over the \cite{Takeuchi2002} prescription.\\
\section{Parameter space exploration} \label{sec_Appendix_ParamSpaceExplore}
\begin{figure*}
\centering
\includegraphics[width=170mm]{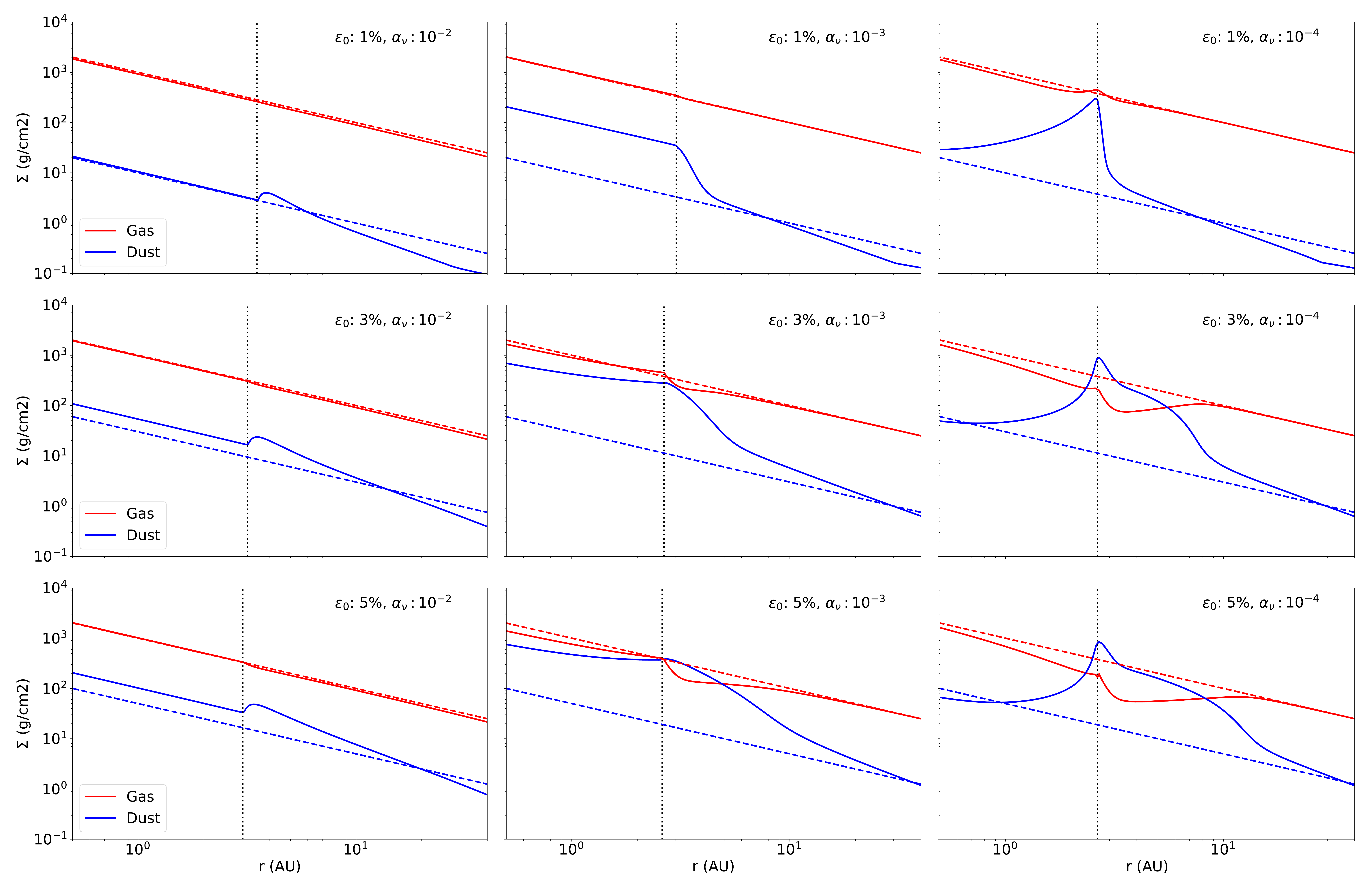}
 \caption{
 Surface density of gas (red) and dust (blue) after $\SI{0.4}{Myr}$ (solid lines) for different values of $\varepsilon_0$ and $\alpha_\nu$. Initial conditions marked with dashed lines. The snowline location marked with dotted lines.
}
 \label{Fig_Appendix_GridDensity}
\end{figure*}
\begin{figure*}
\centering
\includegraphics[width=170mm]{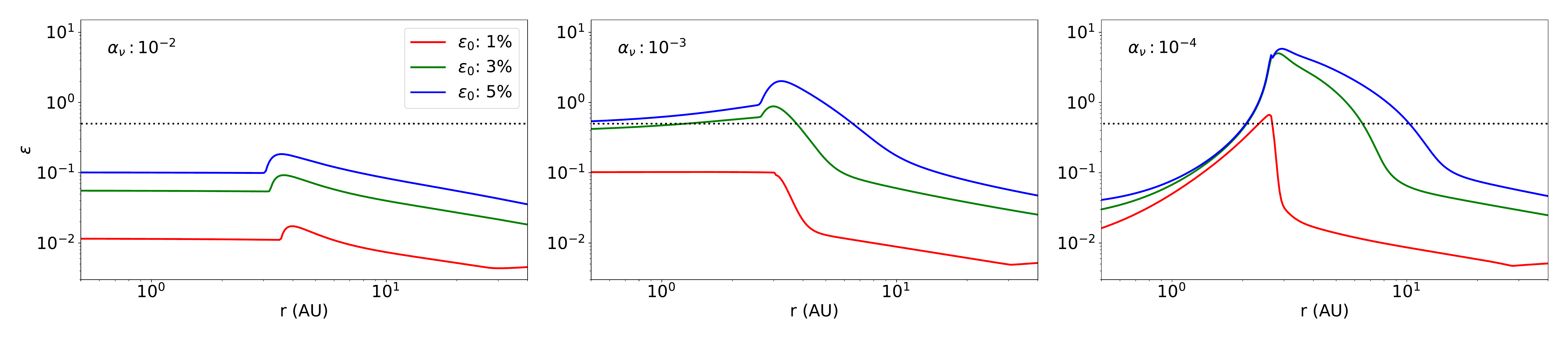}
\includegraphics[width=170mm]{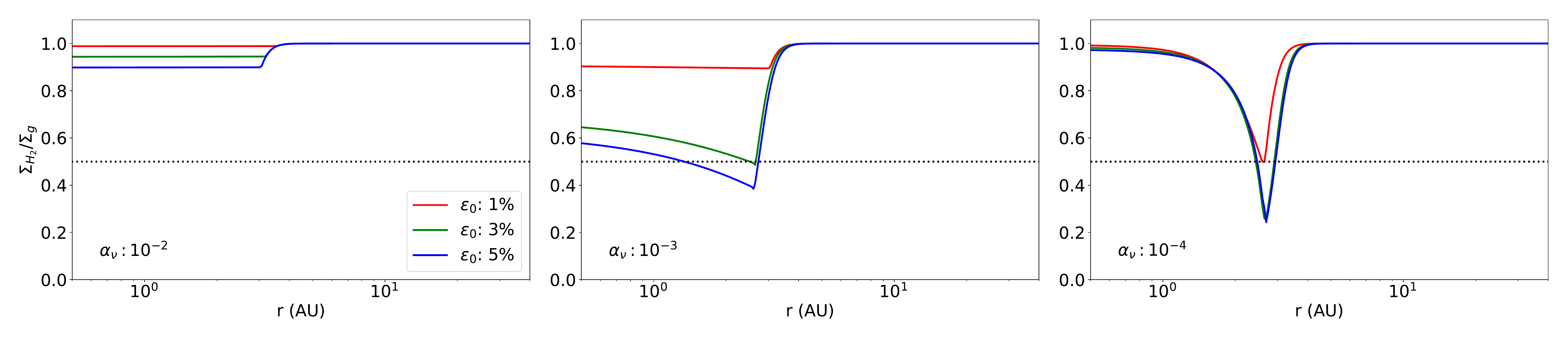}
\includegraphics[width=170mm]{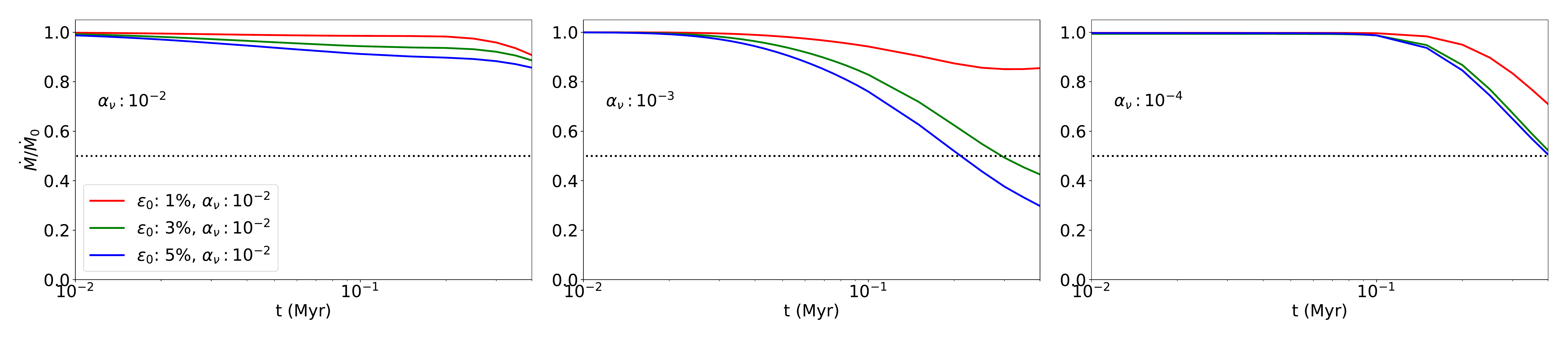}
 \caption{
 \textit{Top Row:} Dust-to-gas ratio radial profile. \textit{Middle Row:} $\textrm{H}_2, \textrm{He}$ mass fraction radial profile. \textit{Bottom Row:} Accretion rate time evolution (divided by the initial steady state accretion $\Dot{M}_0$). All for different values of $\varepsilon_0$ and $\alpha_\nu$. The value of $0.5$ is marked with a dotted line in every plot.
}
 \label{Fig_Appendix_GridRatios}
\end{figure*}
Here we extend the parameter space described in \autoref{TableParam} to show the effect of different turbulent viscosity parameters $\alpha_\nu$ in the global disk evolution (\autoref{Fig_Appendix_GridDensity}), along with the dust-to-gas ratio profile, the $\textrm{H}_2, \textrm{He}$ mass fraction profiles, and the gas accretion rate evolution (\autoref{Fig_Appendix_GridRatios}).\\
We can summarize the plots in a few remarks:
\begin{itemize}
    \item Lower $\alpha_\nu$ values leads to higher dust-to-gas ratios, as the dust is more concentrated towards the snowline and spreads more slowly to the inner boundary.
    \item Higher $\varepsilon_0$ values lead to a stronger perturbation onto the gas surface density, but only if the turbulent viscosity is low enough ($\alpha_\nu \leq \SI{e-3}{}$). 
    \item For an initial $\varepsilon_0 \geq 0.03$ and $\alpha_\nu \leq \SI{e-3}{}$, the dust accumulates both inside and outside the snowline, always reaching dust-to-gas ratios above $\epsilon \geq 0.8$, even if the turbulence is high.
    \item For $\varepsilon_0 \geq 0.03$ and $\alpha_\nu \leq \SI{e-3}{}$, the $\textrm{H}_2, \textrm{He}$ mass fraction is reduced to values between     $0.2 -  0.65$ inside the snowline, though the distribution of water vapor depends on the turbulent viscosity $\alpha_\nu$.
    \item For the case with $\varepsilon_0 = 0.01$ and $\alpha_\nu = \SI{e-4}{}$, the dust concentration is enhanced in a narrow region inside the snowline because of the low viscosity.
    \item The accretion rate can be reduced down to $30\%$ of its initial value depending, on the simulation parameters.
    \item In the high turbulence case ($\alpha_\nu = \SI{e-2}{}$) the dust concentration inside the snowline is reduced to $\epsilon \approx 0.01 - 0.1$, though it can reach to $\epsilon = 0.02 - 0.2$ outside the snowline ($r \approx \SI{3}{}-\SI{4}{au}$) due to the recondensation of water vapor.
\end{itemize}
From these plots we can extract that a global dust-to-gas ratio $\varepsilon_0 \geq 0.03$ and a viscous turbulence $\alpha_\nu \leq \SI{e-3}{}$  are required for the dust back-reaction to perturb the gas surface density, deplete the the inner regions from hydrogen-helium ($\Sigma_{\textrm{H}_2}/\Sigma_\textrm{g} \lesssim 0.5$), reach high concentrations of dust inside and outside the snowline ($\epsilon \gtrsim 0.5$), and finally, to damp the accretion rate ($\Dot{M}/\Dot{M}_0 \approx 0.3 - 0.5$).
\section{Criterion for gap opening through dust back-reaction.}\label{sec_Appendix_BackreactionCriteria}
In this section we will derive a criterion to determine if the dust back-reaction can clear a gap in the gas, based on the transport of angular momentum and the global disk properties.\\
The condition to clear a gap, is that the clearing timescale $t_\textrm{clear}$ must be shorter than the viscous timescale $t_\nu$:
\begin{equation} \label{eq_backreaction_condition_timescale}
t_\textrm{clear} < t_\nu.
\end{equation}
Now we proceed to derive $t_\textrm{clear}$ from the exchange of angular momentum between the dust and gas.\\
The angular momentum of a parcel of material with mass $m$, at a radius $r$, and with orbital velocity $v_K = \sqrt{GM/r}$ is:
\begin{equation} \label{eq_angular_momentum}
    J =  v_K\, r\, m.
\end{equation}
The angular momentum required to transport a ring of material from a radius $r$ to $r_1$ is:
\begin{equation} \label{eq_angular_momentum_ring}
    \stdiff{J} =\left(\sqrt{\frac{r_1}{r_0}} - \sqrt{\frac{r}{r_0}}\right)\, v_0 r_0\, \stdiff{m},
\end{equation}
where $v_0$ is the keplerian velocity at radius $r_0$, and the mass of a gas ring is $\stdiff{m} = 2\pi r \Sigma_\textrm{g} \stdiff{r}$.\\
Assuming that the gas surface density is $\Sigma_\textrm{g} = \Sigma_0 (r_0/r)$, then the total angular momentum required to clear a gap between $r_0$ and $r_1$ is given by:
\begin{equation} \label{eq_angular_momentum_clear}
J_\textrm{clear} = \int \stdiff{J} = 2\pi\, v_0 r_0^2 \Sigma_0 \int_{r_0}^{r_1} \sqrt{\frac{r_1}{r_0}} - \sqrt{\frac{r}{r_0}}\, \stdiff{r}.
\end{equation}
Solving the integral we obtain:
\begin{equation} \label{eq_angular_momentum_clear_solution}
J_\textrm{clear} =  2\pi\, v_0 r_0^3 \Sigma_0 \cdot \left(\frac{1}{3} \left(\frac{r_1}{r_0}\right)^{3/2} - \left(\frac{r_1}{r_0}\right)^{1/2} + \frac{2}{3} \right).
\end{equation}
From \autoref{eq_angular_momentum_ring} we can also infer that the dust drifting from a radius $r_1$ to $r_0$ loses angular momentum (and delivers it to the gas) at a rate of:
\begin{equation}\label{eq_angular_momentum_dust_drift}
    \Dot{J}_\textrm{drift} = v_0 r_0 \Dot{M}_\textrm{d}(r_1) \left(\left(\frac{r_1}{r_0}\right)^{1/2} -1 \right),
\end{equation}
where the accretion rate of dust at $r_1$ is:
\begin{equation} \label{eq_dust_accretion_rate}
    \Dot{M}_\textrm{d}(r_1) = 2 \pi r_1 \Sigma_\textrm{d}(r_1) v_\textrm{d,r} (r_1),
\end{equation}
where we will assume an uniform dust-to-gas ratio, using $\Sigma_\textrm{d} = \varepsilon\, \Sigma_\textrm{g}$.\\ 
Then, considering only the drifting component of the dust velocity (see \autoref{eq_dust_vr}) we obtain that:
\begin{equation} \label{eq_dust_drift_rate_extra}
    v_\textrm{d,r}(r_1) = 2 \textrm{St}\, v_P(r_1) = -  \textrm{St}\, \gamma_P \left(\frac{h}{r}\right)^2 v_0 \left(\frac{r_1}{r_0}\right)^{-1/2}, 
\end{equation}
where have taken the limit of small particles ($\textrm{St} \ll 1$), and expanded the expression for the pressure velocity given in \autoref{eq_pressure_velocity_alter} using $h/r = c_s/v_K$.\\
Now we can write the gap opening timescale as:
\begin{equation} \label{eq_time_clear}
    t_\textrm{clear} = \frac{J_\textrm{clear}}{|\dot{J}_\textrm{drift}|} = C\left(\frac{r_1}{r_0}\right)\, \left(\frac{h}{r}\right)^{-2}  \varepsilon^{-1} \textrm{St}^{-1}  v_0^{-1} r_0,
\end{equation}
with $C(x)$ a function of the ratio $x = r_1/r_0$:
\begin{equation}
    C(x) = \frac{x^2 - 3x + 2 \sqrt{x}}{3\gamma_P (\sqrt{x}- 1)}.
\end{equation}
We apply this formula to our simulations, using a scale height of $h/r = 0.05$, $\gamma_P = 2.75$, and study the time required to clear a gap between $r_0 = \SI{2.5}{au}$ (which is approximately the location of the snowline) and $r_1 = \SI{10}{au}$, and find:
\begin{equation} \label{eq_time_clear_value}
    t_\textrm{clear} \approx \left(\frac{\varepsilon}{0.01}\right)^{-1} \left(\frac{\textrm{St}}{0.03}\right)^{-1} \SI{0.84}{Myr}.
\end{equation}
Now, we only need to compare it with the viscous timescale, which can be understood in this case as the time necessary to close the gap:
\begin{equation} \label{eq_visc_time}
    t_\nu = r_0^2/\nu = \frac{r_0}{v_0 \alpha} \left(\frac{h}{r}\right)^{-2},
\end{equation}
which at the snowline location of $r_0 = \SI{2.5}{au}$ give us a time of:
\begin{equation} \label{eq_visc_time_value}
    t_\nu = \left(\frac{\alpha}{\SI{e-3}{}}\right)^{-1} \SI{0.25}{Myr}.
\end{equation}
To conclude we can derive from \autoref{eq_backreaction_condition_timescale}, \eqref{eq_time_clear} and \eqref{eq_visc_time} a condition on $\alpha$, $\varepsilon$ and $\textrm{St}$ to see if the dust back-reaction can clear a gap:
\begin{equation}
    \frac{\varepsilon \textrm{St}}{\alpha} \gtrsim C(r_1/r_0 = 4) \approx 1,
\end{equation}
where we have taken $r_0 = \SI{2.5}{au}$ and $r_1 = \SI{10}{au}$. Notice that this condition is similar to $2B v_P \gtrsim A v_\nu$ (\autoref{eq_gas_vr}), which indicates if gas motion is locally dominated by the dust back-reaction.\\
To conclude, looking at the values of $t_\textrm{clear}$ and $t_\nu$, it is easy to see why some disks create a gap-like perturbation in \autoref{Fig_Appendix_GridDensity}. The disks with $\alpha = \SI{e-2}{}$ have viscous timescales that are too short in comparison with the clearing timescale by an order of magnitude, while the disks with $\alpha = \SI{e-4}{}$ are easily dominated by the dust back-reaction, provided that dust is delivered for enough time to complete a clearing timescale.
\end{appendix}

\end{document}